%% file: spectrum_fit.tex
\documentstyle[12pt,aasms4]{article}

\input psfig.sty
\input fmacro.tex   

\def\etal{{\sl et~al.}}
\def\eg{{\it e.g.}}
\def\ie{{\it i.e.}}

\def\signal{\vec{s}}
\def\theory{\vec{t}}
\def\yff{{Y_{\rm ff}}}
\def\vsol{{369}}     
\def\vsolerr{{1.5}}  
\def\vgv{{633.9}}     
\def\vgverr{{10}}   
\def\apj{Astrophysical Journal}
\def\prl{Phys. Rev. Let.}

\lefthead{Nordberg \& Smoot}
\righthead{CMB Spectrum Analysis}

\begin{document}

\title{The Cosmic Microwave Background Spectrum: \\  
       an Analysis of Observations \\}

\author{ Henrik~P.~Nordberg\altaffilmark{1} 
     and George~F.~Smoot\altaffilmark{2} } 
\altaffiltext{1}{Lund University, Dept. of Physics, S\"{o}lvegatan 14, 
Lund, Sweden \&
LBNL, Bldg 50-351, University of California, Berkeley CA 94720  } 
\altaffiltext{2}{Lawrence Berkeley National Laboratory 
\& Physics Department, University of California, Berkeley CA 94720 }


\noindent
\begin{abstract}

This work presents a detailed analysis of Cosmic Microwave Background (CMB) 
radiation intensity observations. 
The CMB is a relic of the Big Bang and its study greatly enhances 
our knowledge of cosmology.
This work has led to new values for the  best fit temperature 
$2.7356 \pm 0.0038$~K (95\%\ CL) and 
the speed of our solar system relative to the CMB, 
and limits on the spectral distortion parameters: $\mu$, $y$, and $Y_{\rm ff}$.
These in turn lead to tighter constraints on the thermal history
and allowed energy release in the early universe and 
possible cosmological processes.
These limits are approaching the order of magnitude required
by known processes.
\end{abstract}


\section{Introduction} 

\subsection{The Cosmic Microwave Background Radiation}

Early nucleosynthesis calculations by Gamow, Alpher and Herman 
[\cite{alpher88}] showed that, from the theory of a hot big bang, 
one can infer that a cosmic microwave background radiation
(CMB) will be present today as a result of the primordial fireball. 
The spectrum was predicted to be of black body
form with a temperature of a few Kelvin. 
In 1964 \cite{penzias65} 
discovered the predicted relic radiation 
by measuring a universal ``background noise'' of about 3K. 
This discovery of the cosmic microwave background (CMB)
led to the ascendancy of the hot big bang model of cosmology.

Thermal equilibrium was easily reached in the early universe as a result of
the high rate of interactions between matter and radiation due to the high
density and temperature.
The Planckian distribution is invariant under the universal expansion 
when expressed in dimensionless frequency $x = h \nu / kT = hc / \lambda kT$.
(After decoupling, the photon number is conserved, 
so the occupation number, $n=1/\left(\exp{\frac{h\nu}{kT}}-1\right)$,
and therefore $x = \frac{h\nu}{kT}$ is a constant for each quantum state.) 
Even when the matter and radiation are no longer in good
thermal contact the Planckian shape is preserved.
Therefore we expect 
the CMB spectrum to be Planckian today and, indeed, to a large extent it is.

The CMB spectrum would have a blackbody form if the simple, hot, Big Bang
model is a correct description of the early universe, 
but will be distorted from that form by energy release 
for a redshift $z \lsim3\times10^6$ [\cite{sunyaev80}]. 
Such releases might arise from decay of unstable particles, dissipation of
cosmic turbulence and gravitational waves, breakdown of cosmic strings and
other more exotic transformations. 
The CMB was the dominant energy field after the annihilation of positrons 
and the decoupling of neutrinos until 
$ z \gsim 3\times 10^4$. 
For energy release into the electron-proton plasma, between redshift of 
a few times $10^4$ and $10^6$ [\cite{smoot88}] 
the number of Compton scatterings is sufficient to bring the photons 
into thermal equilibrium with the primordial plasma. 
During this time, bremsstrahlung and other radiative processes do not have 
enough time to add a sufficient amount of photons to create a 
Planckian distribution. 
The resulting distribution is a Bose-Einstein spectrum 
with a chemical potential, $\mu$, that is exponentially attenuated
at low frequencies, $\mu = \mu _0 e^{-2\nu_b /\nu}$ 
(where $\nu_b$ is the frequency at which Compton scattering of photons 
to higher frequencies is balanced by bremsstrahlung creation 
at lower frequencies [\cite{danese82}]).
At redshifts smaller than $\sim10^5$, 
the Compton scattering rate is no longer high
enough to produce a Bose-Einstein spectrum. 
The resulting spectrum has an increased brightness temperature in the far 
Rayleigh-Jeans region due to bremsstrahlung emission by relatively hot 
electrons,
a reduced temperature in the middle Rayleigh-Jeans region where the photons are
depleted by Compton scattering, and a high temperature in the Wien region where 
the Compton-scattered photons have accumulated.

\section{Theory}

\subsection{Spectral Distortions}
\label{thedistortions}

There are three important distortions, Compton ($y$), 
chemical potential ($\mu$) and free-free (or bremsstrahlung, $\yff$).

\subsubsection{Compton Distortion}
Compton scattering ($\gamma e~\rightarrow~\gamma'e'$) of the 
background photons by a hot electron gas, creates spectral distortions
by transferring energy from the electrons to the photons. 
The Compton scattering distortion is characterized by the $u$ parameter
\begin{equation}
	u = -\int_0^z{ {kT_e(z')-kT_\gamma(z') \over m_ec^2 }
		\sigma_Tn_e(z')c{dt \over dz'}dz' }
\end{equation}
where $\sigma_Tn_e(z')cdt$ is the number of interactions and 
$(kT_e(z')-kT_\gamma(z'))/m_ec^2$ is the mean fractional photon-energy
change per collision [Sunyaev \& Zel'dovich 1980]. In the literature 
it is more common to use $y$ 
\begin{equation}
	y = -\int_0^z{ {kT_e(z')\over m_ec^2 }
		\sigma_Tn_e(z')c{dt \over dz'}dz' }
\end{equation}
as a parameter rather than $u$. 
For $T_\gamma\ll T_e$, $y\approx u$.
For $ux^2\ll1$, 
Danese and DeZotti \cite{Danese78} found that
\begin{equation}   
      n_{\rm Compton} \approx {1\over e^x-1}\left\{1+u{xe^x\over e^x-1}\left[
	{x\over\tanh{x/2}}-4\right]\right\}.
\end{equation}	   

For a reasonable number of scatterings, but for $u$ (or $y$) $<< 1$,
each photon is randomly fractionally shifted in energy according 
to a distribution which is Gaussian 
and its energy is increased by the factor $1+u$ (or $1+y$).
Thus a $u$ (or $y$) corresponds to a convolution of Planckian distributions
with mean temperature $(1+u)T_o$.

Compton scattering in effect boosts the photons to a higher frequency.
The resulting spectrum is characterized by a constant decrement in the
Rayleigh-Jeans part of the spectrum,
\begin{equation}
\Delta T_y = 2y T_{\gamma}
\end{equation}
where there are too few photons relative to a blackbody spectrum,
and an exponential rise in temperature
in the Wien region where there are now too many photons.
The magnitude of the distortion is related to the total energy transfer 
[\cite{sunyaev70}]
\begin{equation}
\frac{\Delta {\rm E}}{\rm E} = {\rm e}^{4y} - 1 \approx 4y.
\end{equation}

A Compton $y$ distortion is characteristic of a hot plasma
(e.g. $T_{\rm e} \sim 10^6$ K)
at relatively recent epochs,
$z < 10^5$,
e.g., from a hot intergalactic medium.
Compton scattering alters the photon energy distribution
while conserving photon number.
After many scatterings,
the system will approach statistical (not thermodynamic) equilibrium,
described by the Bose-Einstein distribution

\subsubsection{Bose-Einstein or Chemical Potential $\mu$ Distortion}

As $u$ (or $y$) increases above 1, \ie, after several Compton scatterings, 
the photons and the electrons will have reached statistical equilibrium 
(as opposed to thermodynamic equilibrium) and the photon distribution is a 
Bose-Einstein distribution with a dimensionless chemical potential, 
$\mu$
\begin{equation}
	\label{eq:bose_einstein}
	n_{\mu} = {1\over e^{x+\mu}-1}
\end{equation}
where $x\equiv h\nu/kT$ is the dimensionless frequency
and $\mu$ dimensionless chemical potential is
\begin{equation}
\mu_0 = 1.4 \frac{\Delta {\rm E}}{\rm E}.
\end{equation}

The chemical potential arises from the fact that the number of photons is 
conserved during Compton scattering, 
but the average energy per photon increases.

The equilibrium Bose-Einstein distribution results from the oldest
non-equilibrium processes ($10^5<z<8\times10^6$) [\cite{smoot96}], 
such as the decay of relic particles or primordial inhomogeneities.

Another effect of the hot electrons is Bremsstrahlung (or free-free 
radiation) $$eZ~\rightarrow~e'Z'\gamma.$$ The electrons radiate as 
they are retarded or accelerated in collisions with themselves and 
the other constituents of the primordial plasma, \eg, protons. 
Free-free photons are created at low frequencies, 
and Compton-scattering migrates their energy too slowly, so 
that they thermalize the spectrum to
the electron temperature at low frequencies. 
Taking this effect into account, 
we have a frequency dependent chemical potential
\begin{equation}
	\mu(x)=\mu_0e^{-2x_b/x}
\end{equation}
where $x_b$ is the frequency at which Compton scattering of photons to
higher frequencies is balanced by free-free creation at low frequencies.

Including the free-free emission as well,
the photon occupation number becomes
\begin{equation}
n = \frac{1}{ {\rm e}^{x + \mu(x)} - 1} {\rm e}^{- \yff / x^2 }
        +
        \frac{1 - {\rm e}^{- \yff / x^2}}
             { {\rm e}^{x_{\rm e}} - 1}
\label{eta_mu}
\end{equation}
where $\yff$ is the free-free emissivity/absorptivity coefficient
defined in equation \ref{eq:Yff}.

The resulting spectrum is characterized
by a sharp drop in brightness temperature
with a maximum distortion
\begin{equation}
\Delta T_y \approx 6 T_\gamma \Omega_b^{-7/8} \frac{\Delta E}{E}
\end{equation}
occuring at frequency $2 x_b$
or wavelength
$\lambda \approx 13.6 \Omega_b^{-1}$ cm [\cite{Burigana91}].

\subsubsection{Free-free Distortion}

For very late energy release ($z\ll10^3$),
free-free emission can create rather than erase spectral distortion
in the late universe.
Since the Universe is ionized out to a redshift of at least 5,
the most relevant conditions for a free-free distortion are
a relatively recent reionization $(5 < z < 10^3)$ and
a warm intergalactic medium.  
The free-free distortion arises
because of the lack of Comptonization at recent epochs.

The effect on the present-day CMB spectrum is described by
\begin{equation}
\Delta T_{f\!f} = T_{\gamma}\; \yff / {x^2},
\end{equation}
where $T_{\gamma}$ is the undistorted photon temperature,
$x$ is the dimensionless frequency, and
$Y_{f\!f}/x^2$ is the optical depth to free-free emission:
\begin{equation}
Y_{f\!f}
= \int^z_0 ~\frac{T_e(z') - T_{\gamma}(z') }{ T_e(z') }
\frac{ 8 \pi e^6 h^2 n_e^2 \;g }{ 3 m_e (kT_{\gamma})^3\;
 \sqrt{6\pi\, m_e\, k T_e} }
\;\frac{dt}{dz'} dz'
\label{eq:Yff}
\end{equation}
where $h$ is Planck's constant,
$n_e$ is the electron density and $g$ is the Gaunt
factor [\cite{bartlett91}].

The free-free spectrum is described by the $\yff$ parameter 
\begin{equation}
	\label{eq:free_free}
	n_{\yff}={1\over e^{x/\left(1+\yff/x^2\right)}-1}.
\end{equation}
or, taking into account the possibility of high free-free opacity
at very low frequencies,
\begin{equation}
n_{\yff} =  \frac{ \rm{e}^{-\yff / x^2} }
             { \rm{e}^x - 1 }
  +
        \frac{ 1 - \rm{e}^{-\yff / x^2} }
             {\rm{e}^{x_e} - 1 } .
\label{eta_ff}
\end{equation}

Energy released at different epochs
probes different physical conditions in the early universe
and creates different signatures in the CMB spectrum.
Figure \ref{fig:spectrum} shows the current spectrum observations
and sample spectral distortion
characteristic of each mechanism.
Energy release at recent epochs ($z < 1000$)
will re-ionize the intergalactic medium,
which then cools through free-free emission.
If the gas is hot enough or the release occurs before recombination
($z < 10^5$), Compton scattering of CMB photons
from hot electrons provides the primary cooling mechanism.
Early energy release ($10^5 < z < 10^7$) from relic decay
reaches statistical equilibrium,
characterized by a chemical potential distortion at long wavelengths.


\subsection{The Dipole}
Because the earth orbits the sun, the sun orbits in our galaxy and 
our galaxy moves relative to distant matter, we can have a net velocity
with respect to the background radiation. 
The Doppler effect gives a {\it dipole anisotropy}, which, 
if we convert into thermodynamic temperature, 
measures $3.363 \pm 0.0045$~mK\footnote{Result from this work} 
(95\% CL) in the direction\footnote{Galactic coordinates} 
$(l,b) = (264.31\deg \pm 0.04\deg \pm 0.16\deg, 48.05\deg \pm 0.02\deg 
\pm 0.09\deg$), 
where the first uncertainties are statistical and the second are systematic
[\cite{lineweaver96}].
It implies a speed of our solar system, 
again with respect to the background radiation, of \vsol\ $\pm$ \vsolerr\ km/s 
(95\% CL).
Adding this vector to that of the solar system with respect to the Local Group, 
$v_{\rm \Lsun,LG}=316\pm10$ km/s $(l,b)=(93\pm4\deg,-4\pm4\deg)$ (95\% CL) 
[\cite{aj111}], 
we find that the Local Group of galaxies is heading, at a speed
of \vgv\ $\pm$ \vgverr\ km/s (95\% CL), 
towards the Great Attractor located close to the Andromeda constellation,
\ie\, $(l,b)=(269\deg \pm 4\deg, 28\deg \pm 4\deg$) (95\% CL). 

The Doppler effect change $T$ so that observers
with a velocity $\vec{\beta} \equiv \vec{v}/c$ through a Planckian
radiation field of temperature $T_0$, will measure directionally 
dependent temperatures
\begin{equation}
T_{\rm obs}(\theta)=T_0{\sqrt{(1-\beta^2)}\over(1-\beta\cos\theta)},
\end{equation}
where $\theta$ is the angle between $\vec\beta$
and the direction of observation as measured in the observer's
frame [\cite{peebles93}].


An observer in an isotropic  photon distribution, $n(\nu)$, 
will measure a fractional difference, $\Delta n / n$, between 
the photons per quantum state received in the direction of motion 
and that received in a direction perpendicular to its motion given 
to first order in $\beta$ by:
[\cite{forman70}]
\begin{equation}
	{ \Delta n \over n } = - {d~\ln n\over d~\ln\nu} \beta.
\end{equation}
Thus, we see that dipole anisotropy is frequency dependent and this gives
us an opportunity to detect or set limits on distortion parameters
such as $\mu$, $\yff$ and $y$.

To first order in $\beta$, the dipole anisotropy of the CMB intensity is
\begin{eqnarray}
	\label{eq:dipole_distortion}
	T_d &=& T_{\nu(1+\beta)} - T_\nu  \cr\cr
  &=& {h\nu\over k} \left[   \ln^{-1}\left(1+{1\over n(\nu\left[1+\beta\right])}\right) 
- \ln^{-1}\left(1+{1\over n(\nu)}\right)  \right] \cr\cr
  &\approx& -\frac{h \nu}{k (1+n)} \ln^{-2}\left( 1 + \frac{1}{n} \right)
		\frac{d~\ln n}{d~\ln \nu} \beta.
\end{eqnarray}
For a Planckian spectrum, $n=1/(e^x-1)$, this gives the dipole 
temperature as
\begin{equation}
	T_d\approx T\beta \ \ \left(\equiv T{v\over c}\right).
\end{equation}
Inserting the Bose-Einstein spectrum, eq.~(\ref{eq:bose_einstein}), into 
eq.~(\ref{eq:dipole_distortion}), we get
\begin{equation}
	T_{d,\mu} \approx T\beta \frac{x^2}{(x + \mu)^2}
		\left(1 + \mu  \frac{2x_b}{x^2}\right).
\end{equation}
Similarly, inserting the spectrum with the free-free distortion, 
eq.~(\ref{eq:free_free}), into eq.~(\ref{eq:dipole_distortion}) gives
\begin{equation}
	T_{d,\yff} \approx 3T\beta \frac{\yff}{x^2}.
\end{equation}
The Bose-Einstein and free-free distortions 
are plotted in fig.~\ref{fig:dipole_distortions}.



\section{Observations}
\label{fit:mondip}
\begin{figure}[tb]
\psfig{figure=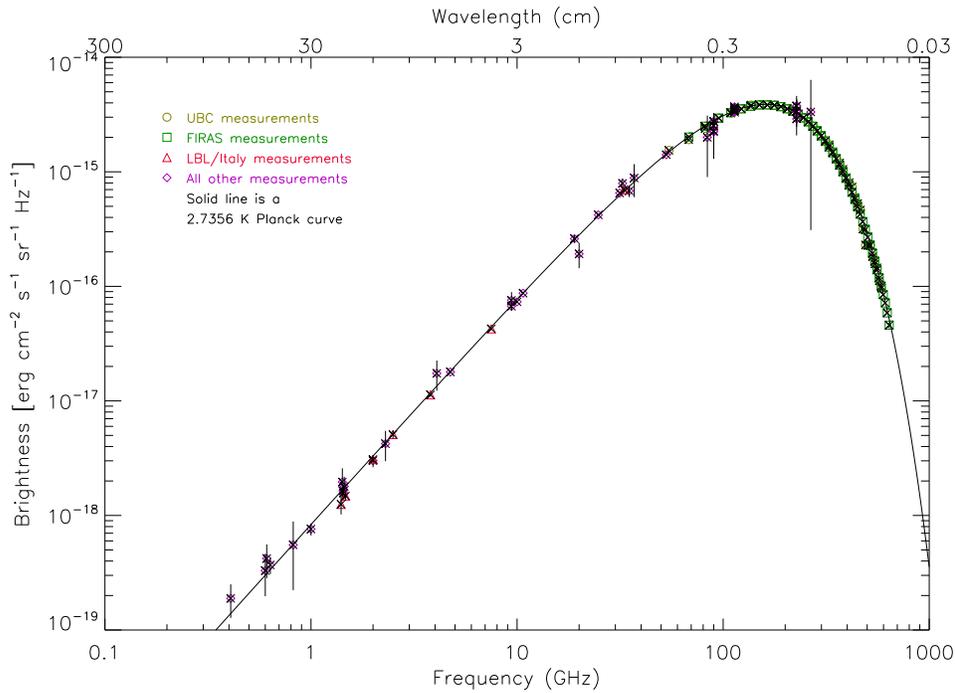,angle=90.,height=5.0in}
\caption{Brightness as a function of frequency of the CMB
\label{fig:spectrum}} 
\end{figure}

\begin{figure}[tb]
\psfig{figure=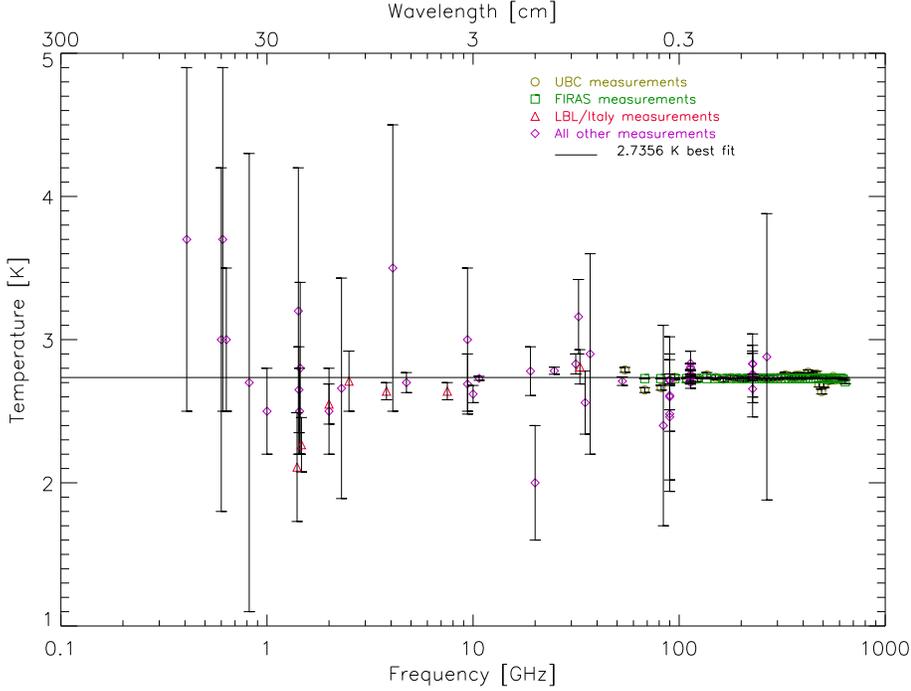,angle=90.,height=5.0in}
\caption{Thermodynamic temperature as a function of frequency of the CMB
\label{fig:tspectrum}} 
\end{figure}

\begin{figure}[tb]
\psfig{figure=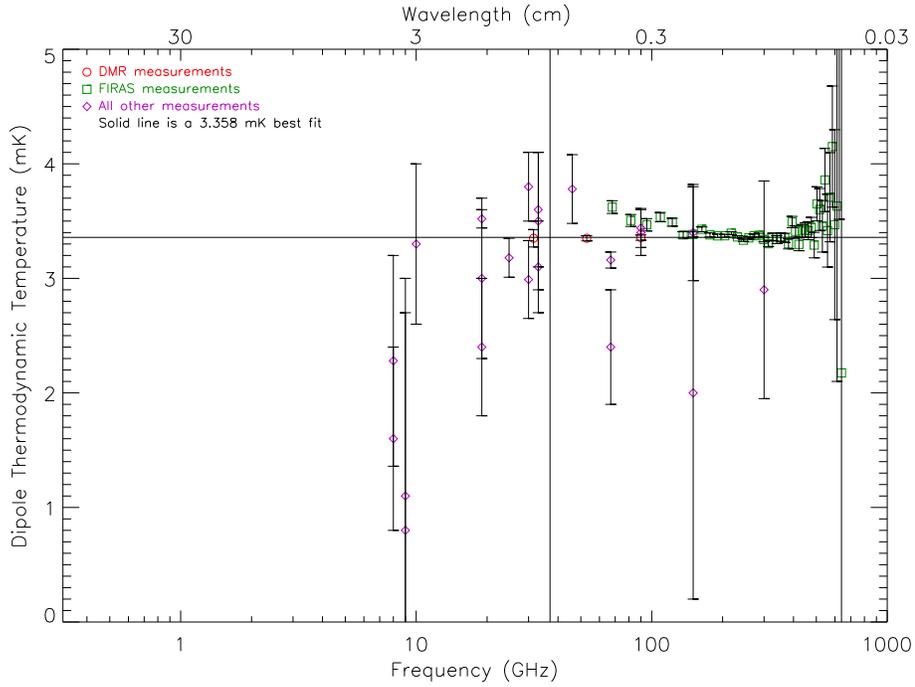,angle=90.,height=5.0in}
\caption{CMB dipole thermodynamic temperature as a function of frequency.
\label{fig:dipole_distortions}}
\end{figure}

We attempted to collect 
all published data, from Penzias's and Wilson's first measurement 
through the recent ground-breaking COBE FIRAS [\cite{fixsen96}]
and COBE DMR [\cite{lineweaver96}] measurements.
All observations are used in this analysis. 
The data used can be found in Tables~\ref{tab:firast} to \ref{tab:mastert2}.
The data are also plotted in Figures~\ref{fig:spectrum} and \ref{fig:tspectrum}
for the temperature measurements and in Figure \ref{fig:dipole_distortions}
for the dipole amplitude.


\begin{figure}[tb]
\psfig{figure=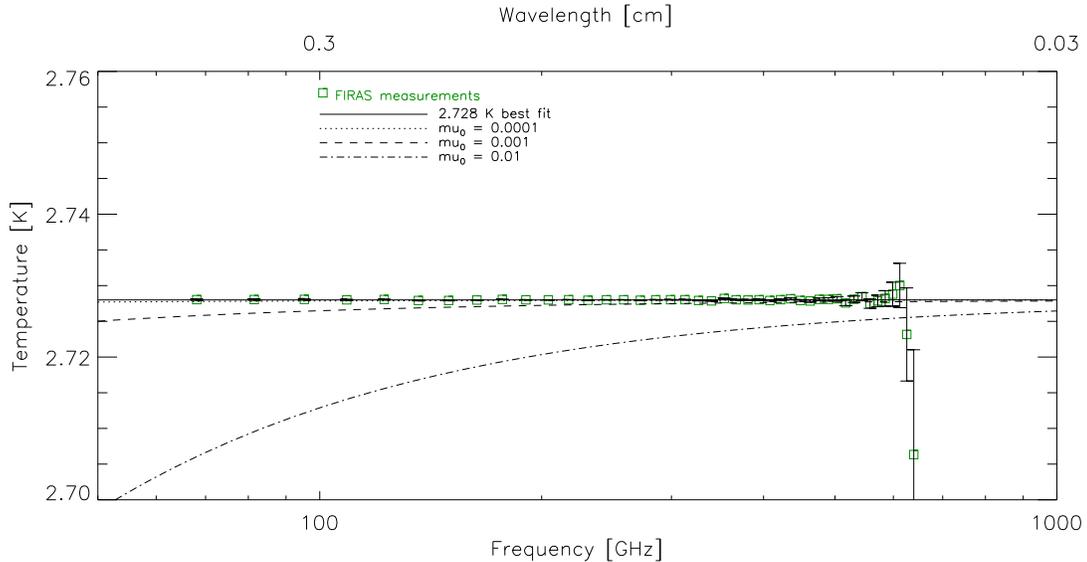}
\caption{Thermodynamic temperature as a function of frequency of the CMB 
monopole.
This is a close-up of Fig. 1 
showing the FIRAS measurements.\label{fig:tspectrumzoom}}
\end{figure}

\begin{figure}[b]
\psfig{figure=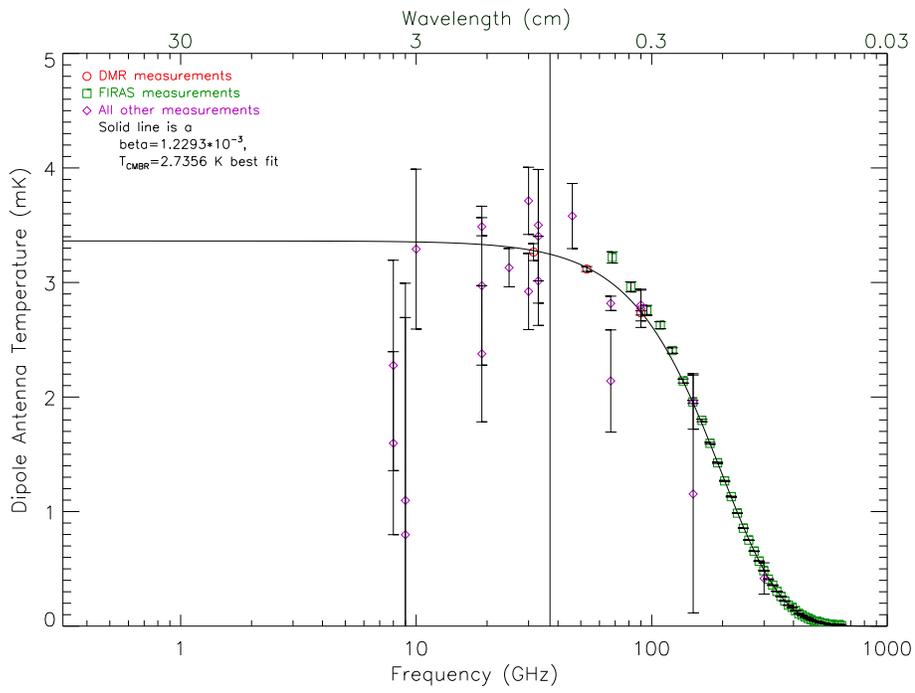,angle=90.,height=5.0in}
\caption{Antenna temperature for the CMB dipole with distortions 
versus frequency.} 
\label{fig:adipole_distortions}
\end{figure}

\begin{figure}[tb]
\psfig{figure=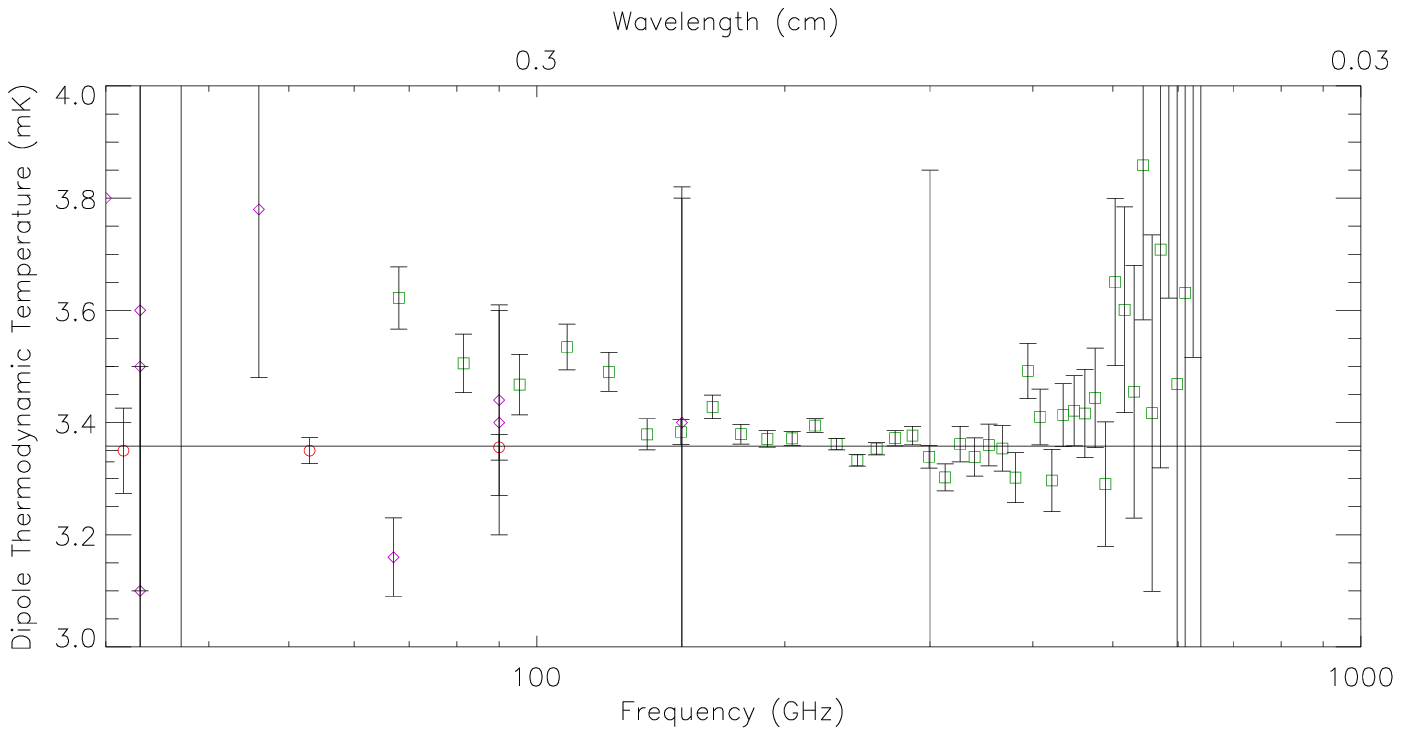}
\caption{CMB dipole thermodynamic temperature as a function of frequency.
This is a close-up of Fig. 5 
Note the position and small errors of the COBE/DMR measurements (circles) 
compared to COBE/FIRAS (squares) and other measurements (diamonds).
\label{fig:dipole_distortions_zoom}}
\end{figure}

We increased the error estimates of a few of the FIRAS and UBC points 
at high frequencies to account for possible systematic errors, 
specifically galactic dust emission removal. 
It is these expanded error estimates that are given in 
Tables~\ref{tab:firast} and \ref{tab:ubct}.


\section{Method}
\label{method}
We fitted the data through finding the minimum $\chi^2$.
Much of the data were reported as individual observations
with a combined statistical and systematic error.
These could in general be treated as independent, uncorrelated data points.
However, much of the data including some of the most significant came
from multiple observations (e.g. FIRAS, UBC, LBL) which share common
systematics and calibration errors.
We accounted for this via using a covariance matrix approach
in calculation of $\chi^2$.
\begin{equation}
 \chi^2\equiv (\signal-\theory)^T {\bf M}^{-1} (\signal-\theory).
\end{equation}
Here ${\bf M}$ is the covariance matrix of the data, $\signal$ the 
signal and $\theory$ the mathematical incarnation of our assumtion. 
\begin{equation}
 {\bf M} \equiv [{\rm Random\;errors}]+[{\rm Correlated\;errors}]
		+[{\rm Theoretical\;errors}]
\end{equation}
Fitting to a theory that is linear in the parameters
is a simple and well developed procedure in linear algebra.
For those cases where it was appropriate, we made linearized fits.
When it was not appropriate we did a full non-linear fitting.
We ran both approaches on a number of cases for comparison of results.

For the non-linear fitting we found that 
the Levenberg-Marquardt method worked well.
We utilized the Levenberg-Marquardt method to iterate to the optimal $\chi^2$.
It combines the Steepest Descent and Grid Search methods into an
algorithm with a fast convergence.

Figure~\ref{fig:tspectrumzoom} and \ref{fig:dipole_distortions} 
show where the current measurements lie and 
where the distortions might be significant. 
As can be seen in the figures, the dipole 
might play a bigger role in determining 
the parameters of the distortions if more measurements were 
available in the region below 10~GHz.


\subsection{Results} 

\begin{table*}
\caption{Nonlinear fit of $T_{\rm 0}$, $\yff$, $\mu_{\rm 0}$, $y$ and 
$\beta$. $\chi^2=259$ with 201 degrees of freedom.}
\begin{center}
\begin{tabular}{llrcll}
$T_0	$&=&$  2.7356		$&$\pm$&$  0.0037         $ K&(95\% CL)\\
$\yff	$&=&$ -1.1\times10^{-5}	$&$\pm$&$  2.3\times10^{-5}$ &(95\% CL)\\
$\mu_0	$&=&$ -3.1\times10^{-5}	$&$\pm$&$  1.2\times10^{-4}$ &(95\% CL)\\
$y	$&=&$  1.6\times10^{-6}	$&$\pm$&$  9.6\times10^{-6}$ &(95\% CL)\\
$\beta	$&=&$  1.2293\times10^{-3} $&$\pm$&$  4.9\times10^{-6}$ &(95\% CL)\\
\end{tabular}
\label{tab:nonlinear}
\end{center}
\begin{center}
Correlation matrix = \hspace{-.15in} 
\begin{math}
\begin{array}{c} \\ T_0 \\ \yff \\ \mu_0 \\ y \\ \beta \end{array} \hspace{-.1in}
\begin{array}{c} \\ \left[ \hspace{-.15in} \begin{array}{c} \\ \\ \\ \\ \\ \end{array} \right. \end{array}
\begin{array}{rrrrr} 
	T_0& \yff& \mu_0&  y &  \beta \\
    1.00&  0.01&  0.05&  0.09&  0.27 \\
    0.01&  1.00& -0.27& -0.18&  0.01 \\
    0.05& -0.27&  1.00&  0.82& -0.01 \\
    0.09& -0.18&  0.82&  1.00&  0.00 \\
    0.27&  0.01& -0.01&  0.00&  1.00 \\
\end{array}
\begin{array}{c} \\ \left. \begin{array}{c} \\ \\ \\ \\ \\ \end{array} \hspace{-.15in} \right] \end{array}
\end{math}
\end{center}
\end{table*}
The results of this work are new limits on the following cosmological 
parameters: \newline
$T_0$ -- the thermodynamic temperature of the background radiation     \newline
$\beta$ -- the quotient of the speed of our solar system to the speed of light, 
$c$ \newline
$\yff$ -- a measure of the effects of Bremsstrahlung, unitless; a.k.a. 
${J_{\rm ff}}$ \newline
y -- a measure of the effects of Compton scattering, unitless	\newline
$\mu_0$ -- describes the Bose-Einstein distortion, unitless.	\newline
The latter three parameters are explained in section \ref{thedistortions}.
The dipole appears because the solar system moves relative to the CMB.

The most complete fit is a nonlinear simultaneous fit of $T_0$, 
$\yff$, $\mu_0$, $y$ and $\beta$, as shown in Table \ref{tab:nonlinear},
making use of all data published for the CMB monopole and dipole. 
The data are dominated by the COBE FIRAS 
[\cite{fixsen96}] and the COBE DMR [\cite{lineweaver96}] measurements 
for the monopole and the dipole respectively.

Table \ref{tab:linear} contains the best-fitted values of $T_0$, 
$\yff$ and $\mu_0$. That fitting was done using both
monopole and dipole data with linearized theory (see section 
\ref{fit:mondip} below). Table \ref{tab:nonlinear} includes $y$ and $\beta$,
as well as the parameters in Table \ref{tab:linear}.
All measurements of the CMB monopole are plotted in 
Figure \ref{fig:spectrum} -- background flux versus frequency, 
Figure \ref{fig:tspectrum} -- temperature versus frequency and 
Figure \ref{fig:tspectrumzoom} -- a close-up of the most interesting part 
of Figure \ref{fig:tspectrum}. 
The dipole is plotted in 
Figure \ref{fig:dipole_distortions} -- dipole temperature versus frequency, 
Figure \ref{fig:dipole_distortions_zoom} -- a close-up of 
Figure \ref{fig:dipole_distortions}
and Figure \ref{fig:adipole_distortions} -- dipole antenna temperature
versus frequency. 
\begin{table*}
\caption{Linear fit with $T_0$, $\yff$, $\mu_0$ and $y$}
\begin{center}
\begin{tabular}{llrcll}
$T_0     $&=&$  2.7356               $&$\pm$&$  0.0038           $ K&(95\% CL)\\
$\yff    $&=&$ -1.1\times10^{-6}     $&$\pm$&$  2.3\times10^{-5} $ &(95\% CL)\\
$\mu_0   $&=&$ -3.0\times10^{-5}     $&$\pm$&$  1.2\times10^{-4} $ &(95\% CL)\\
$y	 $&=&$  1.6\times10^{-6}     $&$\pm$&$  9.6\times10^{-6} $ &(95\% CL)\\
%
%
\end{tabular}
\label{tab:linear}
\end{center}
\begin{center}
 Correlation matrix = \hspace{-.15in} 
\begin{math}
	\begin{array}{c} \\ T_0 \\ \yff \\ \mu_0 \\ y \end{array} \hspace{-.1in}
	\begin{array}{c} \\ \left[ \hspace{-.25in} \begin{array}{c} \\ \\ \\ \\ \end{array} \right. \end{array}
	\begin{array}{rrrr}
  T_0& \yff& \mu_0 & y \\
   1.00 & 0.03 &-0.02 & 0.00 \\
   0.03 & 1.00 &-0.28 &-0.18 \\
  -0.02 &-0.28 & 1.00 & 0.82 \\
   0.00 &-0.18 & 0.82 & 1.00 \\
 	\end{array}
	\begin{array}{c} \\ \left. \begin{array}{c} \\ \\ \\ \\ \end{array} \hspace{-.15in} \right] \end{array}
\end{math}
\end{center}
\end{table*}

\begin{table*}
\caption{Non-linear fit with $T_0$, $\yff$, $\mu_0$ and $y$
to CMB spectrum monopole. {\bf No} dipole data included.}
\begin{center}
\begin{tabular}{llrcll}
$T_0     $&=&$  2.7377 $ & $\pm$ & $  0.0038 $ K &(95\% CL)\\
$\yff    $&=&$ -1.1\times10^{-5}     $&$\pm$&$  2.3\times10^{-5} $ &(95\% CL)\\
$\mu_0   $&=&$ -3.0\times10^{-5}     $&$\pm$&$  1.2\times10^{-4} $ &(95\% CL)\\
$y	 $&=&$  1.6\times10^{-6}     $&$\pm$&$  9.6\times10^{-6} $ &(95\% CL)\\
%
%
\end{tabular}
\label{tab:nlinear}
\end{center}
\begin{center}
 Correlation matrix = \hspace{-.15in} 
\begin{math}
	\begin{array}{c} \\ T_0 \\ \yff \\ \mu_0 \\ y \end{array} \hspace{-.1in}
	\begin{array}{c} \\ \left[ \hspace{-.25in} \begin{array}{c} \\ \\ \\ \\ \end{array} \right. \end{array}
	\begin{array}{rrrr}
  T_0& \yff& \mu_0 & y \\
   1.00 & 0.01 & 0.06 & 0.09 \\
   0.01 & 1.00 &-0.27 &-0.18 \\
   0.06 &-0.27 & 1.00 & 0.82 \\
   0.09 &-0.18 & 0.82 & 1.00 \\
 	\end{array}
	\begin{array}{c} \\ \left. \begin{array}{c} \\ \\ \\ \\ \end{array} \hspace{-.15in} \right] \end{array}
\end{math}
\end{center}
\end{table*}

\section{Analysis of the Cosmic Microwave Background Spectrum}

A novelty of this analysis has been to use the dependence on
frequency for the distortions of the dipole in the fits. 
Using the dipole data in the fit only marginally improved the fit, however.
This is due to the particular frequencies of the dipole measurements. 
Figure~\ref{fig:dipole_distortions} shows the current 
measurements versus frequency and where the distortions are significant. 
As can be seen in Figure~\ref{fig:dipole_distortions}, 
the dipole may play a bigger role in determining the parameters 
of the distortions when more measurements, 
with much greater precision or in the region below 10~GHz, become available. 
Using the dipole does provide
an independent check of the spectrum data.

\begin{figure}[tb]
\psfig{figure=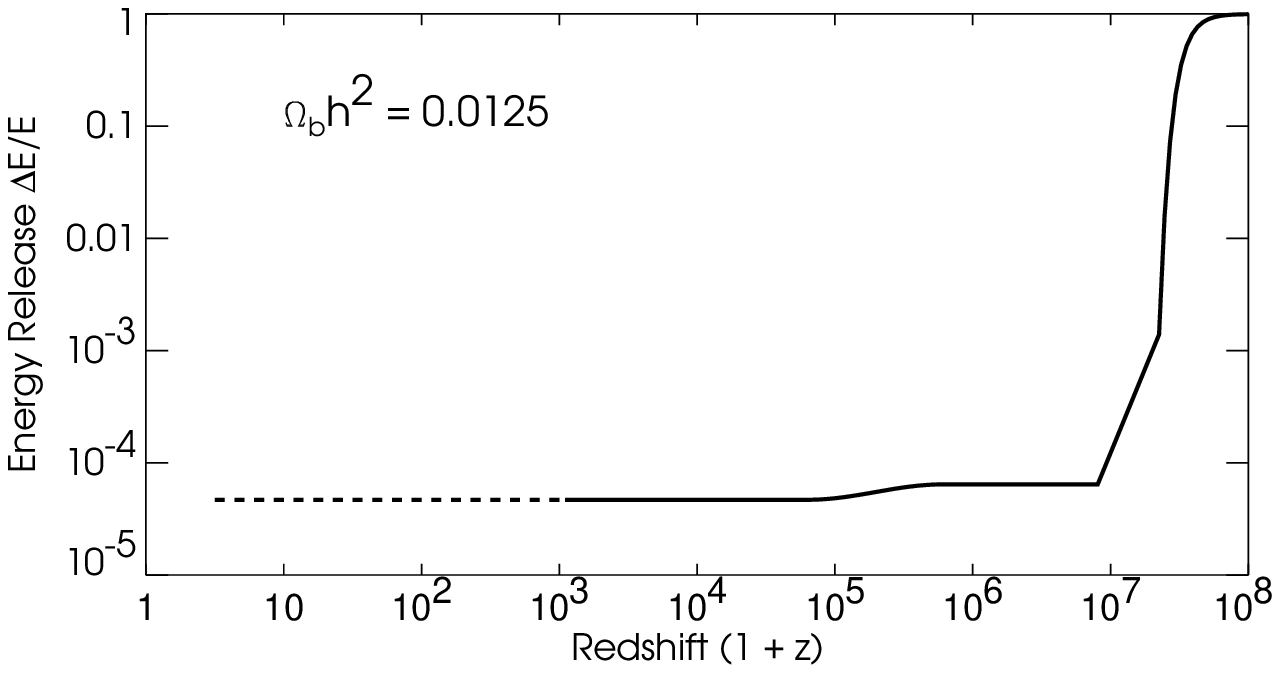,height=3.in}
\caption{Upper limits (95\%\ CL) on fractional energy 
($\Delta E/E_{CBR}$) releases. 
$\Omega_b$ is the baryon number density and $h$ is the hubble expansion rate
per 100 km s$^{-1}$ Mpc$^{-1}$ .
\label{fig:energy_limit}}
\end{figure}

Figure \ref{fig:energy_limit} shows upper limits (95\%\ CL) 
on fractional energy ($\Delta E/E_{CBR}$) releases as set 
by lack of CMB spectral distortions resulting
from processes at different epochs. These can be translated into constraints
on the mass, lifetime and photon branching ratio of unstable relic
particles, with some additional dependence on cosmological parameters
such as $\Omega_{\rm b}$, the baryon number density 
[\cite{hu93}, \cite{wright94}].

\subsection{Some Interpretation of Limits}

The greatest significance of the non-detection of distortions
in the CMB spectrum is the support it gives to
the Hot Big Bang model of cosmology.
A blackbody spectrum is the natural result of a system in thermal equilibrium;
conversely, it is very difficult to superpose a set of non-thermal spectra
to mimic a blackbody to such tight tolerance over so broad a frequency range.
Steady-state models generally create a microwave background
by absorption and subsequent re-emission of starlight by dust.
The re-emitted photons are then redshifted by the expansion of the universe,
so that we observe a superposition of ``shells'' radiating at different epochs.
If the dust opacity were independent of frequency,
the microwave background would be a superposition
of thermal spectra redshifted to different temperatures,
equivalent to a Compton $y$ distortion.
The dust opacity can be tailored to bring the superposition
closer to a blackbody,
but the required (very high) opacities at millimeter wavelengths
conflict with direct observations
of high-redshift galaxies at these wavelengths [\cite{peebles93} p203-6].

The same basic arguments rule out ``plasma universes'' models.

\subsubsection{Stucture Formation by Explosions}
The lack of spectral distortions has implications for structure formation.
Explosive models of structure formation
provide a simple way to sweep (baryonic) matter out of the observed voids.
The resulting shocks, however, heat the baryons and distort the CMB spectrum.
Upper limits on a Compton $y$ distortion
limit the maximum size of explosive voids to
$R_{\rm void} < 2$ Mpc,
much smaller than the observed voids [\cite{levin92}].

\subsubsection{Temperature of the Ionization of the Intergalactic Medium}
The good agreement with a blackbody spectrum also tells us something
about the intergalactic medium (IGM).
A hot ($T_{\rm e} > 10^6$ K) IGM
would provide the high ionization inferred by the lack of a Gunn-Peterson
absorption trough in the spectra of high-redshift quasars.
Thermal bremsstrahlung from the hot gas would provide a natural source
for the observed diffuse X-ray background.
However, a column density large enough to explain the X-ray background
would create an observable Compton $y$ distortion in the CMB spectrum.
Upper limits on the $y$ parameter limit the filling factor of hot gas
to less than $10^{-4}$: 
the diffuse IGM is not very hot, and the X-ray background must come 
from discrete (collapsed) sources [\cite{wright94}].

\subsubsection{Limits on Particle Decay}
Exotic particle decay provides another source for non-zero chemical potential.
Particle physics provides a number of dark matter candidates,
including massive neutrinos, photinos, axions,
or other weakly interacting massive particles (WIMPs).
In most of these models, the current dark matter
consists of the lightest stable member of a family of related particles,
produced by pair creation in the early universe.
Decay of the heavier, unstable members to a photon or charged particle branch
will distort the CMB spectrum provided the particle lifetime is greater
than a year.
Rare decays of quasi-stable particles
(e.g., a small branching ratio for massive neutrino decay
$\nu_{\rm heavy} \rightarrow \nu_{\rm light} + \gamma$)
provide a continuous energy input, also distorting the CMB spectrum.
The size and wavelength of the CMB distortion are dependent upon the
decay mass difference, branching ratio, and lifetime.
Stringent limits on the energy released by exotic particle decay
provides an important input to high-energy theories
including supersymmetry and neutrino physics [\cite{Ellis92}].

\subsubsection{Limits on Antimatter-matter mixing}
In baryon symmetric cosmologies matter-antimatter annihilations give rise
to excessive distortions of the CMB spectrum [\cite{Jones1978}].

\subsubsection{Limits on Primordial Black Hole Evaporation}
Only a very small fraction, $f = M_{planck} / M $, of matter can have formed
black holes in the mass range $10^{11} \leq M \leq 10^{13}$~gm
otherwise their evaporation in the epoch preceding recombination
would have resulted in excessive distortions.
For smaller blackholes the limit is much weaker, since for $M < 10^{11}$~gm,
evaporation would have taken place during the epoch when the photon
spectrum would be completely thermalized.
The constraints follow from the condition that no more than all the entropy
in the universe can come from blackhole evaporation so that
$f < 10^9 M_{planck}/M$.

\subsubsection{Limits on Superconducting Cosmic Strings}
If they are to play an important role in large-scale structure formation,
superconducting cosmic strings would be significant energy sources,
keeping the Universe ionized well past standard recombination.
As a result, the energy input distorts the spectrum of the CMB
by the Sunyaev-Zel'dovich effect. The Compton-$y$ parameter attains a
maximum value in the range of $(1 - 5) \times 10^{-3}$ [\cite{Ostriker87}].
This is well above the observed value.

\subsubsection{Limits on Varying Fundamental Constants}
Noerdlinger [1973] 
pointed out that the intensity of the
Rayleigh-Jeans portion of the CMB spectrum gives the present values
of $kT$, independently of the value of the Planck constant, $h$,
while the wavelength at which the spectrum peaks gives $kT$ in
combination with $h$. That the two temperatures agree
within errors imply that the variation of $h$ must not have exceeded
a few per cent since recombination ($z \sim 1000$).
Likewise a wide variety of $G$-varying cosmologies predict that
the CMB spectrum will follow the standard Planckian formula
multiplied by an epoch-dependent factor, which, in turn, is related
to $G(t)$ [\cite{Narlikar1980}]. The agreement between the brightness
temperature in the Rayleigh-Jeans region and the temperature
determined by the peak location constrain the possible variation
in the gravitational constant $G$.
Likewise one can obtain limits on the variation in the cosmological constant
(energy density of the vacuum) [\cite{Pollock1980}].
The shape of the spectrum also constrains the number of large spatial dimension
(taking into account the possibility of fractal dimensions) to very nearly
three ($\pm 0.02$).

\subsubsection{Limits on Tachyon Coupling Constant}
The existence of particles with space-like four-momenta (``tachyons'')
would make it kinematically possible for a photon to decay into a pair
of tachyons.
If tachyons have any electromagnetic coupling,
photons should have finite lifetimes.
Lorentz invariance dictates that the laboratory lifetime
of a particle scales in proportion to its observed energy ($\gamma$).
The temperature of the CMB does not vary by more than 1\%\
over the range of wavelengths from 0.05 to 1 cm.

Assuming, as would be expected, that the radiation was in thermal equilibrium
when last scattered at a redshift of about 1100, 
the absence of spectral distortions, which would result
from the differential (with energy) photon decay rates
acting over time, argues that the lifetime of microwave photons
must be $\gsim$~40 Hubble times. (Note that the photons begin
with energies a thousand times higher than they currently have and 
thus the effective time for decay is about 40\%\ of the Hubble time.)

This argument [\cite{shapiro91}] sets a more stringent upper limit on the coupling
of tachyons to photons than has previously been reported [\cite{baltay70}].

\subsubsection{Limits on Photon Mass/Oscillations/Paraphotons}
Georgi, Ginsparg, \& Glashow (1983) 
and Okun (1982) 
hypothesized the exisitence of a second
species of photon whose coupling could produce photon masses and 
photon species oscillation.
The oscillations of photon identity are entirely analogous to the
much-discussed neutrino oscillations.
Though the photon mass would have to be much lower than current experimental
limits ($m_\gamma \leq 6 \times 10^{-16}$~eV~$c^{-2}$ at 99.7\%\ CL),
Georgi et al. showed that a photon mass at the 
$m_\gamma = 5 \times 10^{-18} $~eV~$c^{-2}$ could produce distortions
much beyond what is now allowed by the CMB spectrum observations.

The second species of photon would have a very weak interaction with matter
and therefore not have been detected thus far.
This new species of photon would have decoupled much earlier in the big bang
and have a lower temperature than the detected CMB,
since its decoupling would occur earlier.
The intensity of the interacting/oscillating photon species is given by
\begin{equation}
I(\nu) = I_{BB}(\nu, T_1) + [I_{BB}(\nu, T_2) - I_{BB}(\nu, T_1)]
         sin^2(2\phi) sin^2(\pi \nu_0/\nu)
\end{equation}
where $I_{BB}$ is the standard blackbody Planckian spectrum,
and $T_1$ and $T_2$ are the temperature of the two photon species,
$\phi$ is the mixing phase angle, and 
$\nu_0 = 3 \Delta m_\gamma^2 c^4 \tau / 40 \pi^2 \hbar^2 
= 8.2875 \Delta (m_\gamma c^2 / 10^{-18}~{\rm eV})^2~{\rm GHz} $
for $\tau = 1.5 \times 10^{10}$~years
for the age of the Universe (and thus the travel time for the photons).

This manifests itself as an oscillation with frequency of the observed CMB 
intensity or temperature.

This oscillation is very enhanced in the dipole amplitude
and could even change the sign of the dipole
as the dipole amplitude is related to the derivative of the
intensity with respect to frequency and there is a one over frequency ($\nu$)
in the intensity oscillations.
\begin{eqnarray}
\Delta n &=& {\beta \over \nu} \left( {x_1 e^{x_1} \over ( e^{x_1} -1 )^2 } 
+ sin^2(2\phi) \left[  \left( {x_2 e^{x_2} \over ( e^{x_2} -1 )^2 } 
- {x_1 e^{x_1} \over ( e^{x_1} -1 )^2 } \right) sin^2({\pi \nu_0 \over \nu}) 
+ ( { 1 \over e^{x_2} -1} - { 1 \over e^{x_1} -1} )
 sin({2 \pi \nu_0 \over \nu } ) {\pi \nu_0 \over \nu} \right] \right)
\end{eqnarray}
\begin{equation}
{\Delta n \over n} \approx \beta \left( { x_1 e^{x_1} \over e^{x_1} -1 }
+ sin^2(2\phi) \left[  \left( {x_1 e^{x_1} \over e^{x_1} -1}
- {x_2 e^{x_2} \over e^{x_2} -1} \right) sin^2({\pi \nu_0 \over \nu})
- {e^{x_1} - e^{x_2} \over e^{x_2} -1 } sin({2 \pi \nu_0 \over \nu } )
   {\pi \nu_0 \over \nu} \right] \right)
\end{equation}

\begin{figure}
\psfig{figure=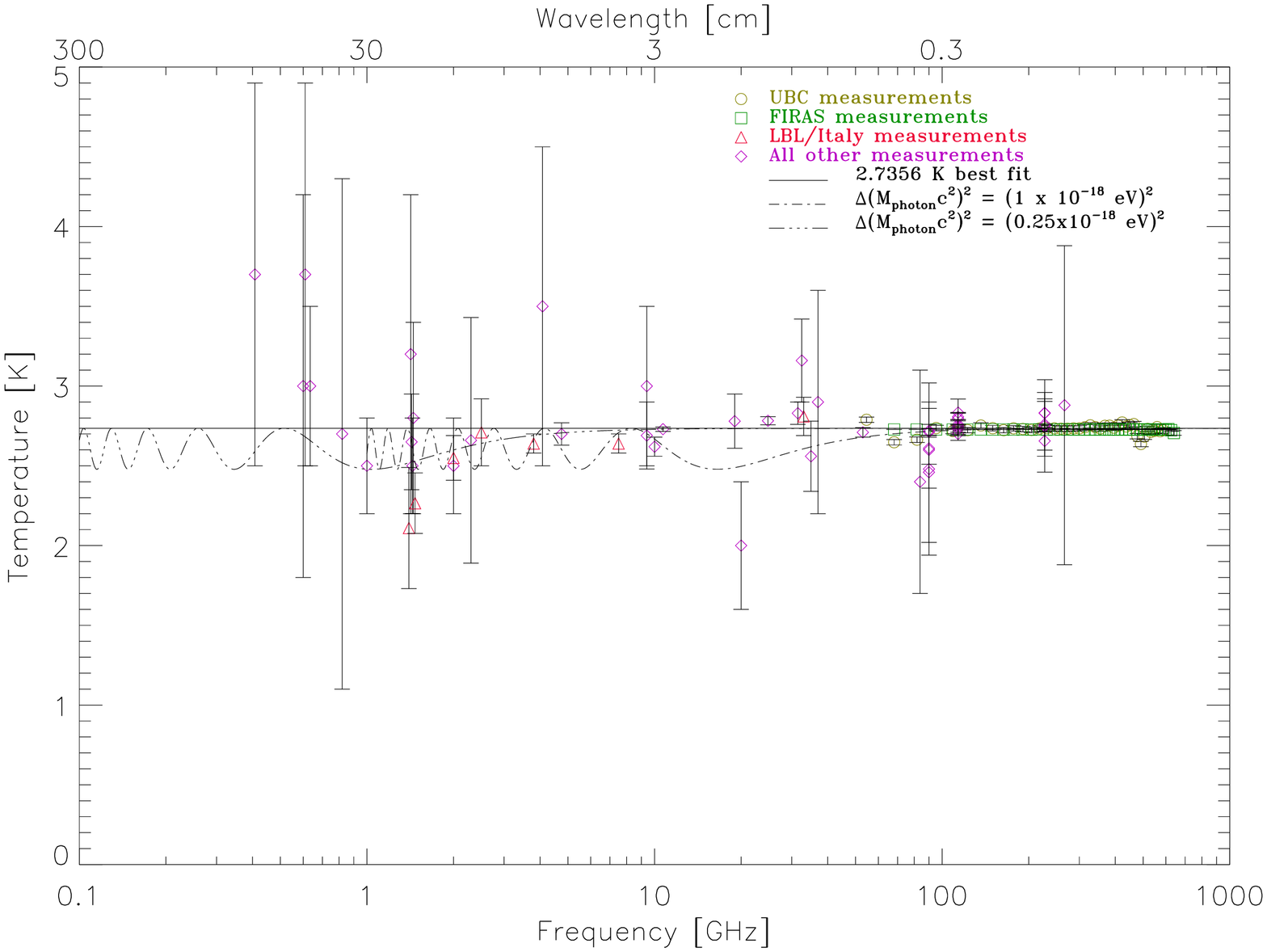,angle=90.,height=5.0in}
\caption{Thermodynamic temperature as a function of frequency of the CMB
for all measurements along with a plot of two photon oscillations
with $\Delta m_\gamma^2 = (1 \times 10^{-18}{\rm eV/c^2})^2$, 
$\Delta m_\gamma^2 = (0.25 \times 10^{-18}{\rm eV/c^2})^2$,
$sin^2(2\phi) = 0.35$ and $T_{2\gamma} = 2.0$~K.
\label{fig:twophotont}} 
\end{figure}

\begin{figure}
\psfig{figure=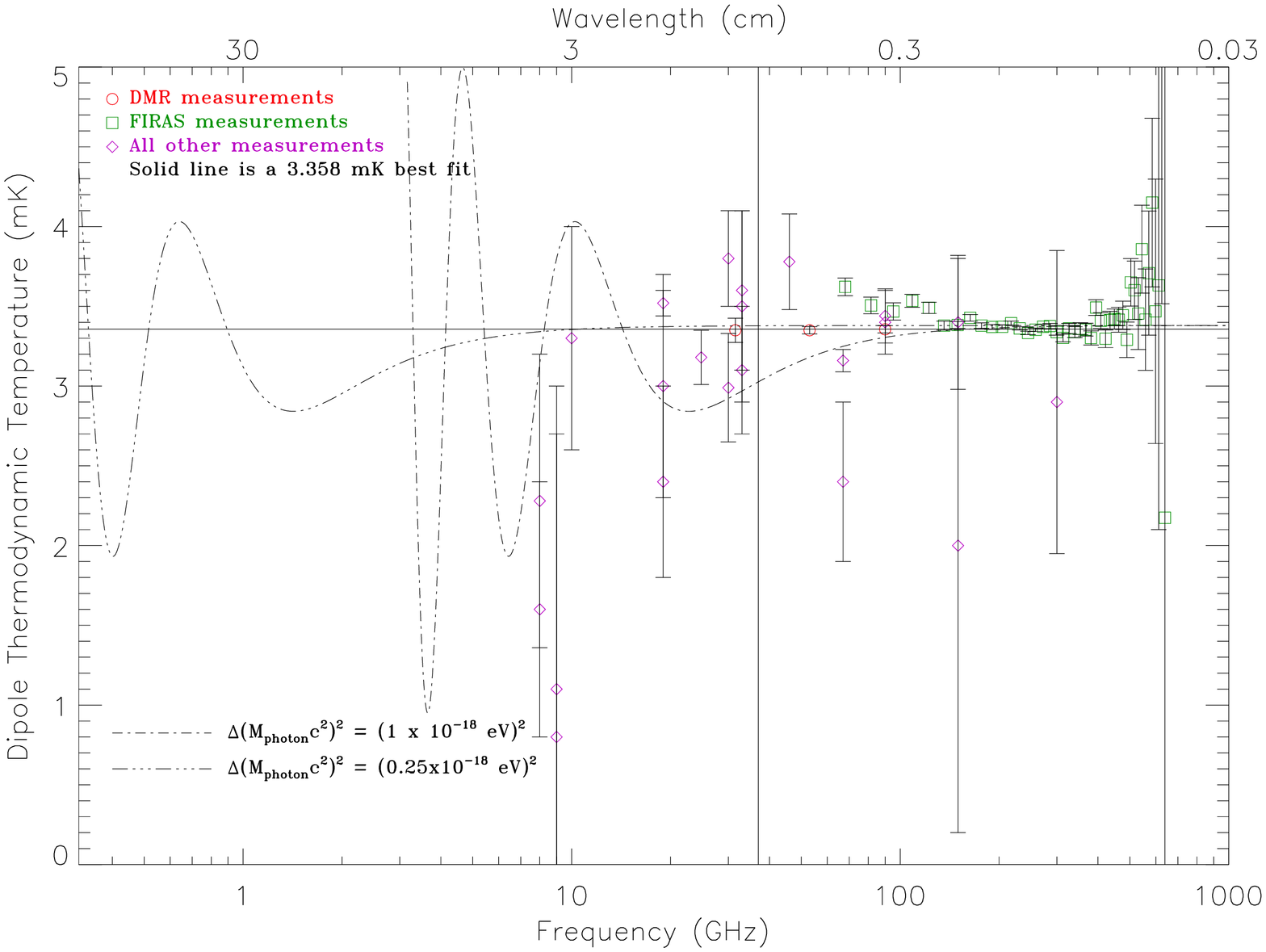,angle=90.,height=5.0in}
\caption{Dipole amplitude in thermodynamic temperature 
as a function of frequency of the CMB for all measurements along 
with a plot of two photon oscillations
with $\Delta m_\gamma^2 = (1 \times 10^{-18}{\rm eV/c^2})^2$, 
$\Delta m_\gamma^2 = (0.25 \times 10^{-18}{\rm eV/c^2})^2$,
$sin^2(2\phi) = 0.35$ and $T_{2\gamma} = 2.0$~K.
\label{fig:twophotond}} 
\end{figure}

Figures \ref{fig:twophotont} and \ref{fig:twophotond} 
show the spectral monopole and dipole distortion 
for the effect of such an oscillation with the parameters: 
$sin^2(2\phi) = 0.35$, $T_1 = 2.74$~K, $T_2 = 2$~K,  and
$\Delta m_\gamma c^2 = 1 \times 10^{-18}$~eV and 
$\Delta m_\gamma c^2 = 0.25 \times 10^{-18}$~eV .


\begin{figure}
\psfig{figure=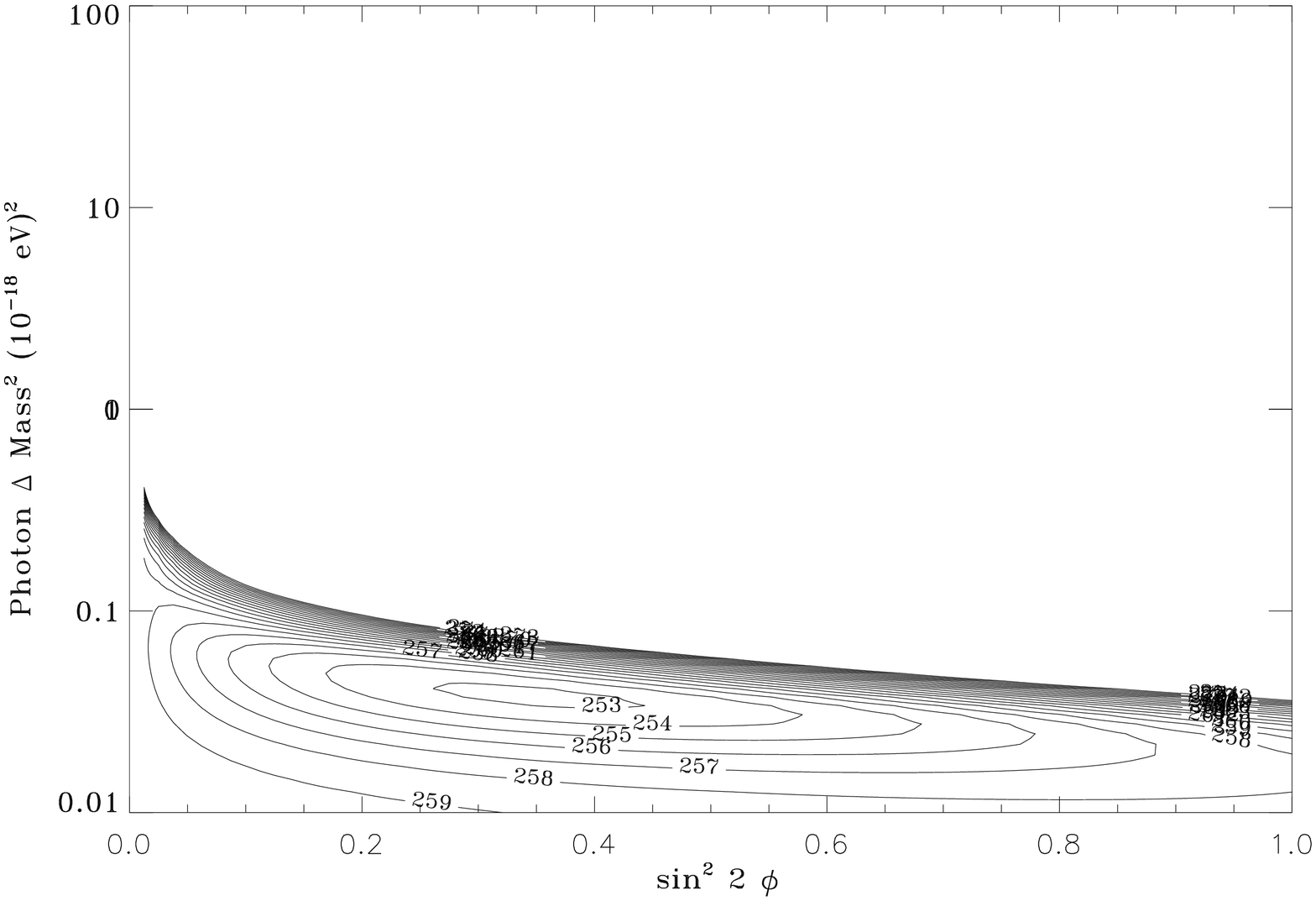,angle=90.,height=5.0in}
\caption{Contour plot of $\chi^2$ on a plane of $sin^2(2\phi)$
versus $\Delta m_\gamma^2$ which is marginalized over the other
parameters.
The $\chi^2$ is for all measurements 
-- temperature and dipole anisotropy amplitude.
\label{fig:twophotonc}} 
\end{figure}

Figure \ref{fig:twophotonc} shows a contour plot of the $\chi^2$ (chi-squared) 
in the $sin^2 2 \phi$ versus $\Delta m_{\gamma}^2$ plane 
of the two-photon oscillating model as fitted to the data.
This plot has minimized $\chi^2$ for the other parameters: 
$T_{CMB}$, $\beta$, and the temperature of the second photon
constrained to be between 0.9 and 2.0 K.
In addition to having a very large region ruled out,
there is a clear minimum in the $\chi^2$ at roughly
$\Delta m_\gamma^2 = (0.25 \times 10^{-18} {\rm eV/c^2})^2$
and $sin^2(2\phi) = 0.35$ in the all data plot.
As can be seen in Figure \ref{fig:twophotont} 
the minimum is due essentially to a limited set of the low frequency
temperature data which are slightly lower than the average.
Such a low photon mass difference produces a lowering of the
predicted temperature at these lower frequencies.
Other than the minimum in $\chi^2$, there is little evidence 
here by the data tracing out the expected form for photon oscillations.
Thus while suggestive, we regard the results thus far as an upper limit
set at a slightly higher $\chi^2$ than normal
and await future measurements. 

Note that going from 
$\Delta m_\gamma^2 = (0.25 \times 10^{-18} {\rm eV/c^2})^2$
to $\Delta m_\gamma^2 = (1 \times 10^{-18} {\rm eV/c^2})^2$
moves the frequency of the oscillations/effect upwards by a factor or 16
and is easily ruled out by both the temperature and dipole amplitude data.
Also know the potential power of low frequency dipole amplitude 
measurements for discriminating photon oscillations.

\subsection{Lower Limits on the Distortions}
\label{lower_limits}

Observations show that the universe is ionized [\cite{haiman97}]. 
The ionized intergalactic medium creates bremsstrahlung and 
sets a lower limit on the free-free distortion to $\yff \sim 2 \times 10^{-6}$. 
This results in an added temperature of $\Delta T_{\rm ff}>0.15$ mK at 2 GHz 
(as low as one might reasonably go in frequency, 
due to synchrotron emision from the galaxy). 
This result only takes into account the observed baryons. 
The actual distortion is bigger,
because this limit did not include bremsstrahlung 
that may have occured at earlier epochs.

A lower limit on the Compton distortion is obtained from measurements of
the Sunyaev-Zeldovich effect [\cite{sunyaev70}]. 
The SZ effect is due to hot ionized plasma component of galactic clusters.
This plasma creates a $y$ for paths that go through the cores 
of galactic clusters. 
Ionized gases in these potential wells Compton-scatter the passing photons. 
Integrating over the full sky, $y$ is of the order of $10^{-6}$ 
[\cite{colafrancesco97}].

Density fluctuations in the early universe oscillate as acoustic waves and
are damped by photon diffusion [\cite{silk68}]. 
The energy dissipated in this manner is added to the radiation field and 
gives rise to a non-zero chemical potential. 
Depending on the nature of the density fluctuations, $\mu$ will
lie between $2\times10^{-8}$ and a few times $10^{-5}$ 
[\cite{daly91}, \cite{hu94}].



\section{Future Experiments}
\label{future_experiments}
Even though no significant distortions to the CMB spectrum have been detected
and since many of the distortions predicted and required by known effects
are not much below current limits,
it is worth considering if future experiments will detect them.
There are two hinderances: (1) the current limits, especially those
set by COBE FIRAS, are quite good and 
(2) there is currently a limited effort in spectrum observations
in part due to (1) and to the high interest in CMB anisotropy
and due to a lack of appreciation 
that the spectral distortions are there to detect.

Since the distortions are frequency dependent an absence of them at 
millimeter and sub-mm wavelengths does {\it not}
imply correspondingly small distortions at centimeter wavelengths.
A better understanding of Galactic emissions, especially at longer 
wavelengths, would help produce tighter limits on the distortion parameters. 
It would be difficult to improve on the COBE measurements without getting 
above the atmosphere. There are, nevertheless,
proposals for new experiments that will measure the CMB. 

ARCADE (Absolute Radiometer for Cosmology, Astrophysics, and Diffuse Emission) 
is a balloon-borne instrument designed to make measurements
in the middle of the spectrum. The first flight will have two channels
at 10 and 30 GHz. Later incarnations will measure the spectrum at
2, 4, 6, 10, 30 and 90 GHz.

The anticipated measurement sensitivity is 1 mK from a balloon, 
limited by the ability to estimate/measure emissions from the 
atmosphere, balloon, flight train, and Earth. 
ARCADE is also a hardware development project 
for the proposed eventual space mission, DIMES. 

The Diffuse Microwave Emission Survey (DIMES) 
has been selected for a mission concept study
for NASA's New Mission Concepts for Astrophysics program 
[\cite{kogut96}, http://ceylon.gsfc.nasa.gov/DIMES/index.html].
DIMES will measure the frequency spectrum of the cosmic microwave
background and diffuse Galactic foregrounds at centimeter wavelengths to
0.1\% precision (0.1 mK). Note that this should detect the free-free distortion,
the lower limit of which was mentioned in section \ref{lower_limits}.
\begin{figure}[tb]
\psfig{figure=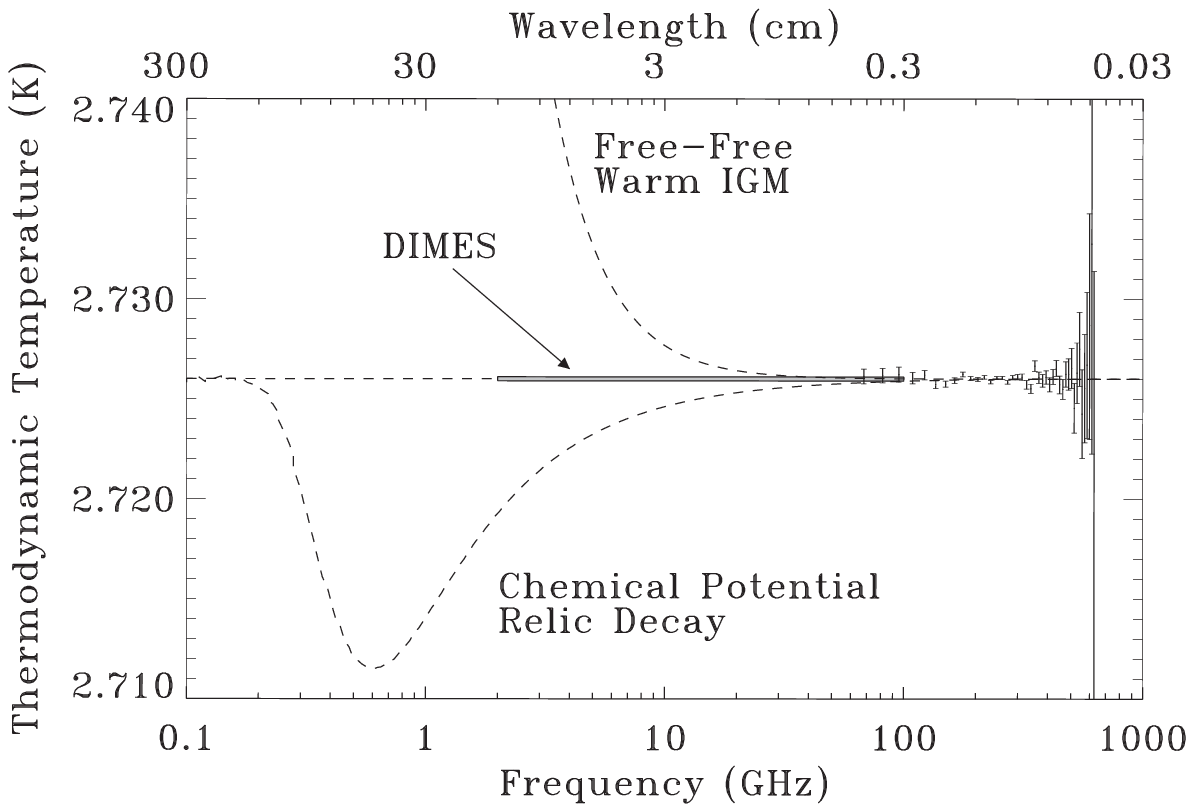}
\caption[DIMES vs FIRAS Sensitivity]
{Current 95\% confidence upper limits to distorted CMB spectra 
appear as dashed lines.
The FIRAS data and DIMES 0.1 mK error box are also shown;
error bars from existing cm-wavelength measurements are larger than
the figure height.
\label{fig:dimes_vs_firas}}
\end{figure}
The FIRAS measurement at sub-mm wavelengths shows no evidence
for Compton heating from a hot IGM (Inter-Galactic Medium).
Since the Compton parameter 
$y \propto n_e T_e$, 
the IGM at high redshift must not be very hot ($T_e \sim 10^5$ K)
or reionization must occur relatively recently ($z_{\rm ion} < 10$).
DIMES will provide a definitive test of these alternatives.
Since the free-free distortion 
$Y_{\rm ff} \propto n_e^2 / \sqrt{T_e}$,  
lowering the electron temperature {\it increases} 
the spectral distortion [\cite{bartlett91}]. 
Figure \ref{fig:dimes_vs_firas} 
shows the limit to $z_{\rm ion}$
that could be established from the combined DIMES and FIRAS spectra,
as a function of the DIMES sensitivity.
A spectral measurement at centimeter wavelengths with 0.1 mK precision 
can detect the free-free signature from the ionized IGM,
allowing direct detection of the onset of hydrogen burning.

ESA's Planck Surveyor mission to map CMB anisotropy
is capable of measuring the CMB dipole with sufficient accuracy
to provide limits at about the current level provided by COBE FIRAS + DMR
in a very independent manner.
This would provide a significant cross-check of these results
and perhaps a small improvement in net results.
It will require dedicated missions such as DIMES to make a really
significant improvement.


\section{Summary}
None of the distortions, $y, \yff$ or $\mu$, are larger than the current
errors in measurement for the monopole and dipole CMB spectra. The monopole
CMB spectrum is very close to a blackbody spectrum. This is strong
evidence of the validity of the hot Big Bang model [\cite{smoot96}]
because there are roughly $10^9$
photons to each baryon in the universe, and it would be extremely difficult
to produce the CMB in astrophysical processes such as the absorption 
and re-emission of starlight by cold dust, 
or the absorption or emission by plasmas, and
still produce such a precise black body spectrum.

The dipole data did not significantly improve on the errors, but were
consistent with and an independent check of the monopole data. 
Future measurements of the dipole could potentially contribute 
to tightening the limits or detecting distortions. 
The Planck Surveyor measurements of the dipole could, in principle, 
determine the distortions to approximately the
same precision as the monopole measurements do now.
Very substantial progress that would surely detect the predicted distortions
and provide us with new information on the thermal history of the Universe 
will require a significant dedicated mission.


\section{Acknowledgments} 
We thank Dr. Alan Kogut for providing us with information on DIMES.
We also thank Dr. S.Y. Frank Chu (LBNL), 
for his insight into numerical programming.

This work supported in part by the DOE
contract No. DE-AC03-76SF00098
through the Lawrence Berkeley National Laboratory.


%
%
\pagebreak
\baselineskip 15 pt
\begin{center}{\bf Dipole References}\end{center}
Bennett, C.L., et al. 1994, \apj, {\bf 436}, 423\newline
Bennett, C.L., et al. 1996, \apj, {\bf 464}, L1-4\newline
Boughn, S.P. et al. 1971, \apj, {\bf 165}, 439\newline
Boughn, S.P. et al. 1981, \apj, {\bf 243}, L113\newline
Cheng, E.S. et al. 1979, \apj, {\bf 232}, L139\newline
Cheng, E.S. 1983 Ph.D. thesis, Princeton Univ.\newline
Conklin, E.K. 1969, Nature, {\bf 222}, 971\newline
Conklin, E.K. 1972, IAU Symposium 44, ed. D.S. Evans (Dordrecht: Reidel), 518\newline
Corey, B.E. 1978, Ph.D. thesis Princeton U.\newline
Corey, B.E. \& Wilkinson D. T., 1976, Bull. Amer. Astron. Soc, {\bf 8}, 351\newline
Cottingham, D.A. 1987, Ph.D. thesis, Princeton Univ.\newline
Fabbri, R., et al. 1980, \prl, 44, 1563, erratum 1980, \prl, {\bf 45}, 401\newline
Fixsen, D.J., Cheng, E.S. \& Wilkinson, D.T. 1983, \prl, {\bf 50}, 620\newline
Fixsen, D.J., et al. 1994, \apj, {\bf 420}, 445\newline
Fixsen, D.J., et al. 1996, \apj, {\bf 486}, 623\newline
Ganga, K., Cheng, E., Meyer, S., Page, L. 1993, \apj, {\bf 410}, L57\newline
Gorenstein, M.V. 1978, Ph.D. thesis, U.C. Berkeley\newline
Halpern, M., et al. 1988 \apj, {\bf 332}, 596\newline
Henry, P.S. 1971, Nature, {\bf 231}, 516\newline
Kogut, A., et al. 1993, \apj, {\bf 419}, 1\newline
Lineweaver, C.H., et al. 1995, Astrophysical Letters and Comm., {\bf 32}, 173-181\newline
Lineweaver, C.H., et al. 1996, \apj, {\bf 470}, 38\newline
Lubin, P.M., Epstein, G.L., Smoot, G.F. 1983, \prl, {\bf 50}, 616\newline
Lubin, P.M. et al. 1985, \apj, {\bf 298}, L1\newline
Meyer, S.S., et al. 1991, \apj, {\bf 371}, L7\newline
Muehler, D. 1976, in Infrared and Submillimeter Ast., ed G.Fazio, D.Reidel, Dordrecht, 63\newline
Partridge, R.B. \& Wilkinson, D. T., 1967, \prl {\bf 18}, 557\newline
Penzias, A.A. \& Wilson, R. W. 1965, \apj, {\bf 142}, 419\newline
Smoot, G.F., Gorenstein, M. V. \& Muller, R. A. 1977, \prl, {\bf 39}, 898\newline
Smoot, G.F. \& Lubin, P. 1979, \apj, {\bf 234}, L83\newline
Smoot, G.F., et al. 1991, \apj, {\bf 371}, L1\newline
Smoot, G.F., et al. 1992, \apj, {\bf 396}, L1\newline
Strukov, I.A., Skulachev, D.P. 1984, Sov. Ast. Lett. {\bf 10}, 3\newline
Strukov, I.A., Skulachev, D.P., Boyarskii, M.N., Tkachev, A.N. 1987, Sov. Ast. Lett. {\bf 13}, 2\newline
Wilkinson, D.T. \& Partridge, R.B. 1969, Partridge quoted in American Scientist, {\bf 57}, 37\newline


\input dipole-table.tex

\smallskip
\input nmaster1.txt
\addtocounter{table}{-1}
\input nmaster2.txt

\input cmb_tables.tex



\end{document}

%% file: fmacro.tex
\def\refitem #1! #2! #3! #4;{\hang\noindent
    \hangindent 20pt\rm #1, \it #2, \bf #3, \rm #4.\par}

\def\folio{\ifnum\pageno=1\nopagenumbers\else\number\pageno\fi}

\def\etal   {{\sl et~al.}}

\def\wisk#1{\ifmmode{#1}\else{$#1$}\fi}

\def\lt     {\wisk{<}}
\def\gt     {\wisk{>}}

\def\lsim   {\wisk{_<\atop^{\sim}}}
\def\gsim   {\wisk{_>\atop^{\sim}}}

\def\Lsun   {\wisk{{\rm L_\odot}}}

\def\deg    {\wisk{^\circ\ }}

\def\singlespace {\smallskipamount=3pt plus1pt minus1pt
                  \medskipamount=6pt plus2pt minus2pt
                  \bigskipamount=12pt plus4pt minus4pt
                  \normalbaselineskip=12pt plus0pt minus0pt
                  \normallineskip=1pt
                  \normallineskiplimit=0pt
                  \jot=3pt
                  {\def\smallskip {\vskip\smallskipamount}}
                  {\def\medskip   {\vskip\medskipamount}}
                  {\def\bigskip   {\vskip\bigskipamount}}
                  {\setbox\strutbox=\hbox{\vrule
                    height8.5pt depth3.5pt width 0pt}}
                  \parskip 0pt
                  \normalbaselines}
\def\doublespace {\smallskipamount=6pt plus2pt minus2pt
                  \medskipamount=12pt plus4pt minus4pt
                  \bigskipamount=24pt plus8pt minus8pt
                  \normalbaselineskip=24pt plus0pt minus0pt
                  \normallineskip=2pt
                  \normallineskiplimit=0pt
                  \jot=6pt
                  {\def\smallskip {\vskip\smallskipamount}}
                  {\def\medskip   {\vskip\medskipamount}}
                  {\def\bigskip   {\vskip\bigskipamount}}
                  {\setbox\strutbox=\hbox{\vrule
                    height17.0pt depth7.0pt width 0pt}}
                  \parskip 12.0pt
                  \normalbaselines}
\def\halfspace {\smallskipamount=6pt plus2pt minus2pt
                  \medskipamount=12pt plus4pt minus4pt
                  \bigskipamount=24pt plus8pt minus8pt
                  \normalbaselineskip=16pt plus0pt minus0pt
                  \normallineskip=2pt
                  \normallineskiplimit=0pt
                  \jot=6pt
                  {\def\smallskip {\vskip\smallskipamount}}
                  {\def\medskip   {\vskip\medskipamount}}
                  {\def\bigskip   {\vskip\bigskipamount}}
                  {\setbox\strutbox=\hbox{\vrule
                    height17.0pt depth7.0pt width 0pt}}
                  \parskip 12.0pt
                  \normalbaselines}

\def\pprintspace {\smallskipamount=4pt plus1pt minus1pt
                  \medskipamount=9pt plus2pt minus2pt
                  \bigskipamount=16pt plus4pt minus4pt
                  \normalbaselineskip=14pt plus0pt minus0pt
                  \normallineskip=1pt
                  \normallineskiplimit=0pt
                  \jot=4pt
                  {\def\smallskip {\vskip\smallskipamount}}
                  {\def\medskip   {\vskip\medskipamount}}
                  {\def\bigskip   {\vskip\bigskipamount}}
                  {\setbox\strutbox=\hbox{\vrule
                   height9.5pt depth4.5pt width 0pt}}
                  \parskip 0pt
                  \normalbaselines}
\def\reidelspace {\smallskipamount=.1667 true in plus4pt minus2pt
                  \medskipamount=.3333 true in plus8pt minus2pt
                  \bigskipamount=13 true pt plus2pt minus2pt
                  \normalbaselineskip=13 true pt plus0pt minus0pt
                  \normallineskip=1 true pt
                  \normallineskiplimit=0 true pt
                  \jot=3pt
                  {\def\smallskip {\vskip\smallskipamount}}
                  {\def\medskip   {\vskip\medskipamount}}
                  {\def\bigskip   {\vskip\bigskipamount}}
                  {\setbox\strutbox=\hbox{\vrule
                    height8.5pt depth3.5pt width 0pt}}
                  \parskip 0pt
                  \normalbaselines}

%% file: dipole-table.tex
\singlespace
\baselineskip 5 pt
\begin{table*}[tb]
\caption{All dipole data used in the fits. Coordinates are galactic.}
\begin{center}
\begin{tabular}{lcccr}
        & Amplitude & Longitude  & Latitude  & Freq\\
	& [mK] & l [deg]& b [deg] & [GHz] \\
\vspace{-.1 in}Penzias \& Wilson (1965) & \lt270 & & & 4   \\
\vspace{-.1 in}Partridge \& Wilkinson (1967)	&   0.8  $\pm$ 2.2  & & & 9   \\  
\vspace{-.1 in}Wilkinson \& Partridge (1969)	&   1.1  $\pm$ 1.6  & & & 9   \\ 
\vspace{-.1 in}Conklin (1969)		&   1.6  $\pm$ 0.8  & 96 $\pm$	30 & 85 $\pm$ 30 & 8  \\
\vspace{-.1 in}Boughn \etal\ (1971)  	&   7.6  $\pm$ 11.6 &              & 	&     37  \\
\vspace{-.1 in}Henry (1971)		&   3.3  $\pm$ 0.7  & 270 $\pm$	30 & 24 $\pm$ 25 & 10 \\
\vspace{-.1 in}Conklin (1972)		&\gt2.28 $\pm$ 0.92 & 195 $\pm$ 30 & 66 $\pm$ 10 & 8  \\
\vspace{-.1 in}Corey \& Wilkinson (1976)&  2.4  $\pm$ 0.6  & 306 $\pm$	28 & 38 $\pm$ 20 & 19 \\
\vspace{-.1 in}Muehler (1976)		&   2.0  $\pm$ 1.8  & 207          & -11         &150 \\
\vspace{-.1 in}Smoot \etal\ (1977) 	&   3.5  $\pm$ 0.6  & 248 $\pm$ 15 & 56 $\pm$ 10 & 33 \\
\vspace{-.1 in}Corey (1978)		&   3.0  $\pm$ 0.7  & 288 $\pm$ 26 & 43 $\pm$ 19 & 19 \\
\vspace{-.1 in}Gorenstein (1978)	&   3.60 $\pm$ 0.5  & 229 $\pm$	11 & 67 $\pm$  8 & 33 \\
\vspace{-.1 in}Cheng \etal\ (1979) 	&   2.99 $\pm$ 0.34 & 287 $\pm$  9 & 61 $\pm$  6 & 30 \\
\vspace{-.1 in}Smoot \& Lubin (1979)	&   3.1  $\pm$ 0.4  & 250.6 $\pm$ 9& 63.2 $\pm$ 6& 33 \\
\vspace{-.1 in}Fabbri \etal\ (1980)	& 2.9  $\pm$ 0.95 & 256.7 $\pm$ 13.8 & 57.4 $\pm$ 7.7 &  300 \\
\vspace{-.1 in}Boughn \etal\ (1981)	&   3.78 $\pm$ 0.30 & 275.4 $\pm$ 3.9  & 46.8	$\pm$  	4.5 &     46  \\   
\vspace{-.1 in}Cheng (1983)		&   3.8  $\pm$ 0.3  & 	               &  &     30  \\
\vspace{-.1 in}Fixsen \etal\ (1983)	&   3.18 $\pm$ 0.17 & 265.7 $\pm$ 3.0  & 47.3	$\pm$    1.5	&     25  \\   
\vspace{-.1 in}Lubin  (1983)		&   3.4  $\pm$ 0.2  &  	    & 	&     90 \\
\vspace{-.1 in}Strukov \etal\ (1984)	&   2.4  $\pm$ 0.5  &  	    & 	&     67  \\	
\vspace{-.1 in}Lubin \etal\ (1985) 	&   3.44 $\pm$ 0.17 & 264.3 $\pm$ 1.9  & 49.2	$\pm$    1.3	&     90  \\   
\vspace{-.1 in}Cottingham (1987)	&   3.52 $\pm$ 0.08 & 272.2 $\pm$ 2.3  & 49.9	$\pm$    1.5	&     19  \\   
\vspace{-.1 in}Strukov \etal\ (1987)	&   3.16 $\pm$ 0.07 & 266.4 $\pm$ 2.3  & 48.5	$\pm$    1.6	&     67  \\   
\vspace{-.1 in}Halpern \etal\ (1988)	&   3.4  $\pm$ 0.42 & 289.5 $\pm$ 4.1  & 38.4	$\pm$    4.8	&     150 \\  
\vspace{-.1 in}Meyer \etal\ (1991) 	&        $\pm$      & 249.9 $\pm$ 4.5  & 47.7	$\pm$    3.0	&     170 \\  
\vspace{-.1 in}Smoot \etal\ (1991) 	&   3.3  $\pm$ 0.1  & 265   $\pm$ 1    & 48 	$\pm$  	1	&     53  \\	
\vspace{-.1 in}Smoot \etal\ (1992) 	&   3.36 $\pm$ 0.1  & 264.7 $\pm$ 0.8  & 48.2	$\pm$    0.5	&     53  \\   
\vspace{-.1 in}Ganga \etal\ (1993) 	&        $\pm$      & 267.0 $\pm$ 1.0  & 49.0	$\pm$    0.7	&     170 \\  
\vspace{-.1 in}Kogut \etal\ (1993) 	&  3.365 $\pm$ 0.027& 264.4 $\pm$ 0.3  & 48.4	$\pm$    0.5	&     53  \\  
\vspace{-.1 in}Fixsen \etal\ (1994)	&  3.347 $\pm$ 0.008& 265.6 $\pm$ 0.75 & 48.3	$\pm$    0.5	&     300 \\  
\vspace{-.1 in}Bennett \etal\ (1994)	&  3.363 $\pm$ 0.024& 264.4 $\pm$ 0.2  & 48.1	$\pm$    0.4	&     53  \\   
\vspace{-.1 in}Bennett \etal\ (1996)	&  3.353 $\pm$ 0.024& 264.26$\pm$ 0.33 & 48.22	$\pm$   0.13	&     53  \\  
\vspace{-.1 in}Fixsen \etal\ (1996)	&  3.372 $\pm$ 0.005& 264.14$\pm$ 0.17 & 48.26	$\pm$   0.16	&     300 \\ 
\vspace{-.1 in}Lineweaver \etal\ (1996)&  3.358 $\pm$ 0.023& 264.31$\pm$ 0.17 & 48.05	$\pm$   0.10	&     53  \\
\end{tabular}
\end{center}
\label{tab:dipole}
\end{table*}

%% file: nmaster1.txt
\begin{table}[htb]
\begin{center}
\caption{Measurements of $T_{\rm CMB}$}
\begin{tabular}{cccll}
\hspace{-.3 in} \small Frequency & \small Wavelength & \small Temperature & \small Type & \small Reference  \\
\hspace{-.3 in} \small (GHz)    & \small (cm)  & \small (K)   &		 &   \\
\hspace{-.3 in} \vspace{-.1 in} \small	0.408 & \small  73.5 & \small $3.7  \pm 1.2$ 	& \small	 Ground (LN)  	& \small	 Howell \& Shakeshaft 1967, Nature, 216, 753.   \\
\hspace{-.3 in} \vspace{-.1 in} \small	0.6   & \small  50   & \small $3.0  \pm 1.2$ 	& \small	 Ground (Term)	& \small	 Sironi et al. 1990, Ap.J., 357, 301.  \\
\hspace{-.3 in} \vspace{-.1 in} \small	0.610 & \small  49.1 & \small $3.7  \pm 1.2$ 	& \small	 Ground (LN)  	& \small	 Howell \& Shakeshaft 1967, Nature, 216, 7  \\
\hspace{-.3 in} \vspace{-.1 in} \small	0.635 & \small  47.2 & \small $3.0  \pm 0.5$ 	& \small	 Ground (LN)  	& \small	 Stankevich et al 1970, Australian J. Phys, 23, 529  \\
\hspace{-.3 in} \vspace{-.1 in} \small	0.820 & \small  36.6 & \small $2.7  \pm 1.6$ 	& \small	 Ground (Term)	& \small	 Sironi et al. 1991, Ap.J., 378, 550.  \\
\hspace{-.3 in} \vspace{-.1 in} \small	1     & \small  30   & \small $2.5  \pm 0.3$ 	& \small	 Ground (LN)  	& \small	 Pelyushenko \& Stankevich 1969, Sov. Astron., 13, 223.  \\
\hspace{-.3 in} \vspace{-.1 in} \small	1.4   & \small  21.3 & \small $2.11 \pm 0.38$ 	& \small	 Ground (CLC)	& \small	 Levin et al. 1988, Ap.J., 334,14 \\
\hspace{-.3 in} \vspace{-.1 in} \small	1.42  & \small  21.2 & \small $3.2  \pm 1.0$ 	& \small	 Ground (Term)	& \small	 Penzias and Wilson 1967, AJ, 72, 315  \\
\hspace{-.3 in} \vspace{-.1 in} \small	1.43  & \small  21   & \small $2.65^{+0.33}_{-0.30}$ 	& \small	 Ground (LN) 	& \small	 Staggs et al. 1996, ApJ, 458, 407\\
\hspace{-.3 in} \vspace{-.1 in} \small	1.44  & \small  20.9 & \small $2.5  \pm 0.3$ 	& \small	 Ground (LN)  	& \small	 Pelyushenko \& Stankevich 1969, Sov. Astron., 13, 223. \\
\hspace{-.3 in} \vspace{-.1 in} \small	1.45  & \small  20.7 & \small $2.8  \pm 0.6$ 	& \small	 Ground (Term)	& \small	 Howell \& Shakeshaft 1966, Nature, 210, 1318. \\
\hspace{-.3 in} \vspace{-.1 in} \small	1.47  & \small  20.4 & \small $2.27 \pm 0.19$ 	& \small	 Ground (CLC)	& \small	 Bensadoun et al. 1993, Ann. NY Acad. Sci, 668, p792-4\\
\hspace{-.3 in} \vspace{-.1 in} \small	2     & \small  15   & \small $2.5  \pm 0.3$ 	& \small	 Ground (LN)  	& \small	 Pelyushenko \& Stankevich 1969, Sov. Astron., 13, 223. \\
\hspace{-.3 in} \vspace{-.1 in} \small	2     & \small  15   & \small $2.55 \pm 0.14$	& \small  Ground (CLC)	& \small  Bersanelli \etal, 1994, Ap.J., 424, 517\\
\hspace{-.3 in} \vspace{-.1 in} \small	2.3   & \small  13.1 & \small $2.66 \pm 0.77$ 	& \small	 Ground (Term)	& \small	 Otoshi \& Stelzreid 1975, IEEE Trans on Inst \& Meas, 24, 174. \\
\hspace{-.3 in} \vspace{-.1 in} \small	2.5   & \small  12   & \small $2.71 \pm 0.21$ 	& \small	 Ground (CLC) 	& \small	 Sironi et al. 1991, Ap. J., 378, 550. \\
\hspace{-.3 in} \vspace{-.1 in} \small	3.8   & \small  7.9  & \small $2.64 \pm 0.06$ 	& \small	 Ground (CLC) 	& \small	 De Amici et al. 1991, Ap.J., 381, 341. \\
\hspace{-.3 in} \vspace{-.1 in} \small	4.08  & \small  7.35 & \small $3.5  \pm 1.0 $ 	& \small	 Ground (Term)	& \small	 Penzias \& Wilson 1965, Ap.J., 142, 419. \\
\hspace{-.3 in} \vspace{-.1 in} \small	4.75  & \small  6.3  & \small $2.70 \pm 0.07$ 	& \small	 Ground (CLC) 	& \small	 Mandolesi et al. 1986, Ap.J., 310, 561. \\
\hspace{-.3 in} \vspace{-.1 in} \small	7.5   & \small  4.0  & \small $2.60 \pm 0.07$ 	& \small	 Ground (CLC) 	& \small	 Kogut et al. 1988, Ap.J., 355, 102\\
\hspace{-.3 in} \vspace{-.1 in} \small	7.5   & \small  4.0  & \small $2.64 \pm 0.06$ 	& \small	 Ground (CLC) 	& \small	 Levin et al. 1992, Ap.J., 396, 3 \\
\hspace{-.3 in} \vspace{-.1 in} \small	9.4   & \small  3.2  & \small $3.0  \pm 0.5 $ 	& \small	 Ground (Term)	& \small	 Roll \& Wilkinson 1966, Phys. Rev. Lett., 16, 405. \\
\hspace{-.3 in} \vspace{-.1 in} \small	9.4   & \small  3.2  & \small $2.69^{+0.16}_{-0.21}$ 	& \small	 Ground (CLC)	& \small	 Stokes et al. 1967, Phys. Rev. Lett., 19, 1199. \\
\hspace{-.3 in} \vspace{-.1 in} \small	10    & \small  3.0  & \small $2.62 \pm 0.06$ 	& \small	 Ground (CLC) 	& \small	 Kogut et al. 1990, Ap.J., 355, 102. \\
\hspace{-.3 in} \vspace{-.1 in} \small	10.7  & \small  2.8  & \small $2.730\pm 0.014$ 	& \small	 Balloon (LHe) 	& \small	 Staggs et al. 1996, Ap.J., 473, L1\\ 
\hspace{-.3 in} \vspace{-.1 in} \small	19.0  & \small  1.58 & \small $2.78^{+0.12}_{-0.17}$ 	& \small	 Ground (CLC)	& \small	 Stokes et al. 1967, Phys. Rev. Lett., 19, 1199. \\
\hspace{-.3 in} \vspace{-.1 in} \small	20    & \small  1.5  & \small $2.0   \pm 0.4 $ 	& \small	 Ground (CLC) 	& \small	 Welch et al. 1967, Phys. Rev. Lett, 18, 1068. \\
\hspace{-.3 in} \vspace{-.1 in} \small	24.8  & \small  1.2  & \small $2.783 \pm 0.025$ 	& \small	 Balloon    	& \small	 Johnson \& Wilkinson 1987, Ap.J. Lett, 313, L1. \\
\end{tabular}
\end{center}
\end{table}

%% file: nmaster2.txt
\begin{table}[htb]
\begin{center}
\caption{Measurements of $T_{\rm CMB}$ {\it continued}}
\begin{tabular}{cccll}
\hspace{-.3 in} \small Frequency \small & Wavelength & \small Temperature & \small Type & \small Reference\\
\hspace{-.3 in} \small (GHz)    & \small (cm)     & \small (K)       &		 &     \\
\hspace{-.3 in} \vspace{-.1 in}  \small	31.5  & \small	0.95  & \small $2.83  \pm 0.07$ 	& \small	 COBE/DMR		& \small	 Kogut et al. 1996, Ap.J, 470, 653\\
\hspace{-.3 in} \vspace{-.1 in}  \small	32.5  & \small	0.924 & \small $3.16  \pm 0.26$ 	& \small	 Ground (CLC)	& \small	 Ewing et al. 1967, Phys. Rev. Lett, 19, 1251. \\
\hspace{-.3 in} \vspace{-.1 in}  \small	33.0  & \small	0.909 & \small $2.81  \pm 0.12$ 	& \small	 Ground (CLC)	& \small	 De Amici et al. 1985, Ap.J., 298, 710. \\
\hspace{-.3 in} \vspace{-.1 in}  \small	35.0  & \small	0.856 & \small $2.56^{+0.17}_{-0.22}$ 	& \small	 Ground (CLC)	& \small	 Wilkinson 1967, Phys. Rev. Lett., 19, 1195. \\
\hspace{-.3 in} \vspace{-.1 in}  \small	37    & \small	0.82  & \small $2.9   \pm 0.7 $ 	& \small	 Ground (LN)	& \small	 Puzanov et al. 1968, Sov. Astr., 11, 905. \\
\hspace{-.3 in} \vspace{-.1 in}  \small	53    & \small	0.57  & \small $2.71  \pm 0.03$ 	& \small	 COBE/DMR		& \small	 Kogut et al. 1996, Ap.J, 470, 653\\
\hspace{-.3 in} \vspace{-.1 in}  \small	83.8  & \small	0.358 & \small $2.4   \pm 0.7 $ 	& \small	 Ground (LN)	& \small	 Kislyakov et al. 1971, Sov. Ast., 15, 29. \\
\hspace{-.3 in} \vspace{-.1 in}  \small	90    & \small	0.33  & \small $2.46^{+0.40}_{-0.44}$ 	& \small	 Ground (CLC)	& \small	  Boynton et al. 1968, Phys. Rev. Lett., 21, 462. \\
\hspace{-.3 in} \vspace{-.1 in}  \small	90    & \small	0.33  & \small $2.61  \pm 0.25$ 	& \small	 Ground (CLC)	& \small	 Millea et al. 1971, Phys. Rev. Lett., 26, 919. \\
\hspace{-.3 in} \vspace{-.1 in}  \small	90    & \small	0.33  & \small $2.48  \pm 0.54$ 	& \small	 Plane (Term)	& \small	 Boynton \& Stokes 1974, Nature, 247, 528. \\
\hspace{-.3 in} \vspace{-.1 in}  \small	90    & \small	0.33  & \small $2.60  \pm 0.09$ 	& \small	 Ground (CLC)	& \small	 Bersanelli et al. 1989, Ap.J., 339, 632. \\
\hspace{-.3 in} \vspace{-.1 in}  \small	90    & \small	0.33  & \small $2.712 \pm 0.020$ 	& \small	 Ground (CLC)	& \small	 Schuster et al. UC Berkeley PhD Thesis \\
\hspace{-.3 in} \vspace{-.1 in}  \small	90    & \small	0.33  & \small $2.72  \pm 0.04$ 	& \small	 COBE/DMR		& \small	 Kogut et al. 1996, Ap.J, 470, 653\\
\hspace{-.3 in} \vspace{-.1 in}  \small	90.3  & \small	0.332 & \small $<2.97			$ 	& \small	  Balloon  	& \small	 Bernstein et al. 1990, Ap.J., 362, 107. \\
\hspace{-.3 in} \vspace{-.1 in}  \small	113.6 & \small	0.264 & \small $2.70  \pm 0.04$ 	& \small	 CN (z Per)	& \small	 Meyer \& Jura 1985, Ap.J., 297, 119. \\
\hspace{-.3 in} \vspace{-.1 in}  \small	113.6 & \small	0.264 & \small $2.74  \pm 0.05$ 	& \small	  CN (z Oph)	& \small	 Crane et al. 1986, Ap.J., 309, 12. \\
\hspace{-.3 in} \vspace{-.1 in}  \small	113.6 & \small	0.264 & \small $2.76  \pm 0.07	$ 	& \small	 CN (HD 21483)	& \small	 Meyer et al. 1989, Ap.J. Lett, 343, L1. \\
\hspace{-.3 in} \vspace{-.1 in}  \small	113.6 & \small	0.264 & \small $2.796^{+0.014}_{-0.039}$ 	& \small	 CN (z Oph)	& \small	 Crane et al. 1989, Ap.J., 346, 136. \\
\hspace{-.3 in} \vspace{-.1 in}  \small	113.6 & \small	0.264 & \small $2.75  \pm 0.04$ 	& \small	 CN (z Per)	& \small	 Kaiser \& Wright 1990, Ap.J. Lett, 356, L1. \\
\hspace{-.3 in} \vspace{-.1 in}  \small	113.6 & \small	0.264 & \small $2.834 \pm 0.085$ 	& \small	  CN (HD 154368)	& \small	 Palazzi et al. 1990, Ap.J., 357, 14. \\
\hspace{-.3 in} \vspace{-.1 in}  \small	113.6 & \small	0.264 & \small $2.807 \pm 0.025$ 	& \small	 CN (16 stars)	& \small	 Palazzi et al. 1992, Ap.J., 398, 53. \\
\hspace{-.3 in} \vspace{-.1 in}  \small	113.6 & \small	0.264 & \small $2.279^{+0.023}_{-0.031}$ 	& \small	 CN (5 stars)	& \small	 Roth et al. 1993, Ap.J., 413, L67. \\
\hspace{-.3 in} \vspace{-.1 in}  \small	154.8 & \small	0.194 & \small $<3.02			 $ 	& \small	 Balloon 	& \small	 Bernstein et al. 1990, Ap.J., 362, 107. \\
\hspace{-.3 in} \vspace{-.1 in}  \small	195.0 & \small	0.154 & \small $<2.91			 $ 	& \small	 Balloon 	& \small	 Bernstein et al. 1990, Ap.J., 362, 107. \\
\hspace{-.3 in} \vspace{-.1 in}  \small	227.3 & \small	0.132 & \small $2.656 \pm 0.057$ 	& \small	 CN (5 stars)	& \small	 Roth et al. 1993, Ap.J., 413, L67. \\
\hspace{-.3 in} \vspace{-.1 in}  \small	227.3 & \small	0.132 & \small $2.76  \pm 0.20$ 	& \small	 CN (z Per)	& \small	 Meyer \& Jura 1985, Ap.J., 297, 119. \\
\hspace{-.3 in} \vspace{-.1 in}  \small	227.3 & \small	0.132 & \small $2.75^{+0.24}_{-0.29}$ 	& \small	 CN (z Oph)	& \small	 Crane et al. 1986, Ap.J., 309, 822. \\
\hspace{-.3 in} \vspace{-.1 in}  \small	227.3 & \small	0.132 & \small $2.83  \pm 0.09$ 	& \small	 CN (HD 21483)	& \small	  Meyer et al. 1989, Ap.J. Lett, 343, L1. \\
\hspace{-.3 in} \vspace{-.1 in}  \small	227.3 & \small	0.132 & \small $2.832 \pm 0.072$ 	& \small	  CN (HD 154368)	& \small	 Palazzi et al. 1990, Ap.J., 357, 14.  \\
\hspace{-.3 in} \vspace{-.1 in}  \small	266.4 & \small	0.113 & \small $<2.88			$ 	& \small	 Balloon  	& \small	 Bernstein et al. 1990, Ap.J., 362, 107. \\
\hspace{-.3 in} \vspace{-.1 in}  \small See Tab. \ref{tab:firast} & \small See Tab. \ref{tab:firast}& \small	 $2.728	\pm 0.002$ 	& \small	 COBE/FIRAS	& \small	 Fixsen et al. 1996, Ap.J., 486, 623. \\
\hspace{-.3 in} \vspace{-.1 in}  \small	300   & \small	0.1   & \small $2.736	\pm 0.017$ 	& \small	 Rocket  	& \small	 Gush et al. 1990, PRL, 65, 537., See also Tab. \ref{tab:ubct} \\
\end{tabular}
\end{center}
\end{table}

%% file: cmb_tables.tex
\begin{table*}
\caption{FIRAS Data;\ \protect{\cite{fixsen96}}}
\begin{center}
\begin{tabular}{rrrr}
Frequency & Temperature & Upper Error & Lower Error\\
GHz & K & K & K\\ 
\hline
\input firast1.tex
\end{tabular}
\end{center}
\label{tab:firast}
\end{table*}

\addtocounter{table}{-1} 
\begin{table*}
\caption{FIRAS Data; \protect{\cite{fixsen96}}\ -- {\it continued}}
\begin{center}
\begin{tabular}{rrrr}
Frequency & Temperature & Upper Error & Lower Error\\
GHz & K & K & K\\
\hline
\input firast2.tex
\end{tabular}
\end{center}
\label{tab:firast2}
\end{table*}

\begin{table*}
\caption{UBC COBRA Rocket Data;\ \protect{\cite{gush90}}}
\begin{center}
\begin{tabular}{rrrr}
Frequency & Temperature & Upper Error & Lower Error\\
GHz & K & K & K\\ 
\hline
\input ubct1.tex
\end{tabular}
\end{center}
\label{tab:ubct}
\end{table*}

\addtocounter{table}{-1} 
\begin{table*}
\caption{UBC COBRA Rocket Data; \protect{\cite{gush90}}\ -- {\it continued}}
\begin{center}
\begin{tabular}{rrrr}
Frequency & Temperature & Upper Error & Lower Error\\
GHz & K & K & K\\
\hline
\input ubct2.tex
\end{tabular}
\end{center}
\label{tab:ubct2}
\end{table*}

\begin{table*}
\caption{All CMB measurements except FIRAS and UBC COBRA}
\begin{center}
\begin{tabular}{rrrr}
Frequency & Temperature & Upper Error & Lower Error\\
GHz & K & K & K\\ 
\hline
\input mastert1.tex
\end{tabular}
\end{center}
\label{tab:mastert}
\end{table*}

\addtocounter{table}{-1}
\begin{table*}
\caption{All CMB measurements except FIRAS and UBC COBRA -- {\it continued}}
\begin{center}
\begin{tabular}{rrrr}
Frequency & Temperature & Upper Error & Lower Error\\
GHz & K & K & K\\
\hline
\input mastert2.tex
\end{tabular}
\end{center}
\label{tab:mastert2}
\end{table*}

%% file: firast1.tex
$    68.1$&$  2.72804$&$   .00011$&$   .00011$\\
$    81.5$&$  2.72805$&$   .00011$&$   .00011$\\
$    95.3$&$  2.72807$&$   .00011$&$   .00011$\\
$   108.8$&$  2.72801$&$   .00009$&$   .00009$\\
$   122.3$&$  2.72806$&$   .00007$&$   .00007$\\
$   136.1$&$  2.72792$&$   .00006$&$   .00006$\\
$   149.6$&$  2.72792$&$   .00005$&$   .00005$\\
$   163.4$&$  2.72798$&$   .00004$&$   .00004$\\
$   176.9$&$  2.72807$&$   .00004$&$   .00004$\\
$   190.4$&$  2.72801$&$   .00003$&$   .00003$\\
$   204.2$&$  2.72800$&$   .00003$&$   .00003$\\
$   217.6$&$  2.72803$&$   .00002$&$   .00002$\\
$   231.1$&$  2.72795$&$   .00002$&$   .00002$\\
$   244.9$&$  2.72802$&$   .00002$&$   .00002$\\
$   258.4$&$  2.72802$&$   .00002$&$   .00002$\\
$   272.2$&$  2.72795$&$   .00003$&$   .00003$\\
$   285.7$&$  2.72802$&$   .00003$&$   .00003$\\
$   299.2$&$  2.72803$&$   .00004$&$   .00004$\\
$   313.0$&$  2.72803$&$   .00005$&$   .00005$\\
$   326.5$&$  2.72792$&$   .00006$&$   .00006$\\
$   340.0$&$  2.72786$&$   .00007$&$   .00007$\\
$   353.8$&$  2.72820$&$   .00008$&$   .00008$\\
$   367.2$&$  2.72802$&$   .00008$&$   .00008$\\

%% file: firast2.tex
$   381.0$&$  2.72798$&$   .00009$&$   .00009$\\
$   394.5$&$  2.72803$&$   .00010$&$   .00010$\\
$   408.0$&$  2.72792$&$   .00010$&$   .00010$\\
$   421.8$&$  2.72803$&$   .00011$&$   .00011$\\
$   435.3$&$  2.72816$&$   .00012$&$   .00012$\\
$   448.8$&$  2.72792$&$   .00013$&$   .00013$\\
$   462.6$&$  2.72785$&$   .00015$&$   .00015$\\
$   476.1$&$  2.72807$&$   .00019$&$   .00019$\\
$   489.9$&$  2.72807$&$   .00023$&$   .00023$\\
$   503.4$&$  2.72816$&$   .00030$&$   .00030$\\
$   516.8$&$  2.72757$&$   .00037$&$   .00037$\\
$   530.6$&$  2.72809$&$   .00045$&$   .00045$\\
$   544.1$&$  2.72845$&$   .00055$&$   .00055$\\
$   557.9$&$  2.72748$&$   .00066$&$   .00066$\\
$   571.4$&$  2.72786$&$   .00080$&$   .00080$\\
$   584.9$&$  2.72821$&$   .00108$&$   .00108$\\
$   598.7$&$  2.72880$&$   .00168$&$   .00168$\\
$   612.2$&$  2.73002$&$   .00311$&$   .00311$\\
$   625.7$&$  2.72316$&$   .00652$&$   .00652$\\
$   639.5$&$  2.70634$&$   .01468$&$   .01468$\\

%% file: ubct1.tex
$    54.6$&$    2.789$&$     .100$&$     .100$\\
$    68.1$&$    2.648$&$     .100$&$     .100$\\
$    81.8$&$    2.664$&$     .100$&$     .100$\\
$    95.3$&$    2.737$&$     .010$&$     .010$\\
$   109.1$&$    2.718$&$     .010$&$     .010$\\
$   122.6$&$    2.724$&$     .010$&$     .010$\\
$   136.4$&$    2.753$&$     .010$&$     .010$\\
$   149.9$&$    2.736$&$     .010$&$     .010$\\
$   163.7$&$    2.724$&$     .010$&$     .010$\\
$   177.2$&$    2.735$&$     .010$&$     .010$\\
$   191.0$&$    2.731$&$     .010$&$     .010$\\
$   204.5$&$    2.725$&$     .010$&$     .010$\\
$   218.2$&$    2.734$&$     .010$&$     .010$\\
$   231.7$&$    2.737$&$     .010$&$     .010$\\
$   245.5$&$    2.735$&$     .010$&$     .010$\\
$   259.0$&$    2.733$&$     .010$&$     .010$\\
$   272.8$&$    2.733$&$     .010$&$     .010$\\
$   286.3$&$    2.735$&$     .010$&$     .010$\\
$   300.1$&$    2.735$&$     .010$&$     .010$\\
$   313.6$&$    2.742$&$     .010$&$     .010$\\
$   327.4$&$    2.754$&$     .010$&$     .010$\\
$   340.9$&$    2.743$&$     .010$&$     .010$\\

%% file: ubct2.tex
$   354.7$&$    2.734$&$     .010$&$     .010$\\
$   368.1$&$    2.751$&$     .010$&$     .010$\\
$   381.9$&$    2.752$&$     .010$&$     .010$\\
$   395.4$&$    2.739$&$     .010$&$     .010$\\
$   409.2$&$    2.752$&$     .010$&$     .010$\\
$   422.7$&$    2.772$&$     .010$&$     .010$\\
$   436.5$&$    2.747$&$     .010$&$     .010$\\
$   450.0$&$    2.755$&$     .010$&$     .010$\\
$   463.8$&$    2.762$&$     .010$&$     .010$\\
$   477.3$&$    2.686$&$     .100$&$     .100$\\
$   491.1$&$    2.637$&$     .100$&$     .100$\\
$   504.6$&$    2.683$&$     .100$&$     .100$\\
$   518.3$&$    2.732$&$     .100$&$     .100$\\
$   531.8$&$    2.713$&$     .100$&$     .100$\\
$   545.6$&$    2.719$&$     .100$&$     .100$\\
$   559.1$&$    2.743$&$     .100$&$     .100$\\
$   572.9$&$    2.717$&$     .100$&$     .100$\\
$   586.4$&$    2.723$&$     .100$&$     .100$\\

%% file: mastert1.tex
\vspace{-.1 in} $   0.408$&$ 3.700$&$ 1.200$&$ 1.200$\\
\vspace{-.1 in} $   0.600$&$ 3.000$&$ 1.200$&$ 1.200$\\
\vspace{-.1 in} $   0.610$&$ 3.700$&$ 1.200$&$ 1.200$\\
\vspace{-.1 in} $   0.635$&$ 3.000$&$  .500$&$  .500$\\
\vspace{-.1 in} $   0.820$&$ 2.700$&$ 1.600$&$ 1.600$\\
\vspace{-.1 in} $   1.000$&$ 2.500$&$  .300$&$  .300$\\
\vspace{-.1 in} $   1.400$&$ 2.110$&$  .380$&$	.380$\\
\vspace{-.1 in} $   1.420$&$ 3.200$&$ 1.000$&$ 1.000$\\
\vspace{-.1 in} $   1.430$&$ 2.650$&$  .330$&$  .300$\\
\vspace{-.1 in} $   1.440$&$ 2.500$&$  .300$&$  .300$\\
\vspace{-.1 in} $   1.450$&$ 2.800$&$  .600$&$  .600$\\
\vspace{-.1 in} $   1.470$&$ 2.266$&$  .190$&$  .190$\\
\vspace{-.1 in} $   2.000$&$ 2.500$&$  .300$&$  .300$\\
\vspace{-.1 in} $   2.000$&$ 2.550$&$  .140$&$  .140$\\
\vspace{-.1 in} $   2.300$&$ 2.660$&$  .770$&$  .770$\\
\vspace{-.1 in} $   2.500$&$ 2.710$&$  .210$&$  .210$\\
\vspace{-.1 in} $   3.800$&$ 2.640$&$  .060$&$  .060$\\
\vspace{-.1 in} $   4.080$&$ 3.500$&$ 1.000$&$ 1.000$\\
\vspace{-.1 in} $   4.750$&$ 2.700$&$  .070$&$  .070$\\
\vspace{-.1 in} $   7.500$&$ 2.640$&$  .060$&$  .060$\\
\vspace{-.1 in} $   9.400$&$ 3.000$&$  .500$&$  .500$\\
\vspace{-.1 in} $   9.400$&$ 2.690$&$  .160$&$  .210$\\
\vspace{-.1 in} $  10.000$&$ 2.620$&$  .060$&$  .060$\\
\vspace{-.1 in} $  10.700$&$ 2.730$&$  .014$&$  .014$\\
\vspace{-.1 in} $  19.000$&$ 2.780$&$  .120$&$  .170$\\
\vspace{-.1 in} $  20.000$&$ 2.000$&$  .400$&$  .400$\\
\vspace{-.1 in} $  24.800$&$ 2.783$&$  .025$&$  .025$\\
\vspace{-.1 in} $  31.500$&$ 2.830$&$  .070$&$  .070$\\

%% file: mastert2.tex
\vspace{-.1 in} $  32.500$&$ 3.160$&$  .260$&$  .260$\\
\vspace{-.1 in} $  33.000$&$ 2.810$&$  .120$&$  .120$\\
\vspace{-.1 in} $  35.000$&$ 2.560$&$  .170$&$  .220$\\
\vspace{-.1 in} $  37.000$&$ 2.900$&$  .700$&$  .700$\\
\vspace{-.1 in} $  53.000$&$ 2.710$&$  .030$&$  .030$\\
\vspace{-.1 in} $  83.800$&$ 2.400$&$  .700$&$  .700$\\
\vspace{-.1 in} $  90.000$&$ 2.460$&$  .400$&$  .440$\\
\vspace{-.1 in} $  90.000$&$ 2.610$&$  .250$&$  .250$\\
\vspace{-.1 in} $  90.000$&$ 2.480$&$  .540$&$  .540$\\
\vspace{-.1 in} $  90.000$&$ 2.600$&$  .090$&$  .090$\\
\vspace{-.1 in} $  90.000$&$ 2.712$&$  .020$&$  .020$\\
\vspace{-.1 in} $  90.000$&$ 2.720$&$  .040$&$  .040$\\
\vspace{-.1 in} $  90.300$&$ < 2.97$&$		$&$		$\\	
\vspace{-.1 in} $ 113.600$&$ 2.700$&$  .040$&$  .040$\\
\vspace{-.1 in} $ 113.600$&$ 2.740$&$  .050$&$  .050$\\
\vspace{-.1 in} $ 113.600$&$ 2.760$&$  .070$&$  .070$\\
\vspace{-.1 in} $ 113.600$&$ 2.796$&$  .014$&$  .039$\\
\vspace{-.1 in} $ 113.600$&$ 2.750$&$  .040$&$  .040$\\
\vspace{-.1 in} $ 113.600$&$ 2.834$&$  .085$&$  .085$\\
\vspace{-.1 in} $ 113.600$&$ 2.807$&$  .025$&$  .025$\\
\vspace{-.1 in} $ 113.600$&$ 2.729$&$  .023$&$  .031$\\
\vspace{-.1 in} $ 227.300$&$ 2.760$&$  .200$&$  .200$\\
\vspace{-.1 in} $ 227.300$&$ 2.750$&$  .240$&$  .290$\\
\vspace{-.1 in} $ 227.300$&$ 2.830$&$  .090$&$  .090$\\
\vspace{-.1 in} $ 227.300$&$ 2.832$&$  .072$&$  .072$\\
\vspace{-.1 in} $ 227.300$&$ 2.656$&$  .057$&$  .057$\\
\vspace{-.1 in} $ 266.400$&$ 2.88$&$   1.0$&$	1.0$\\